\documentclass[12pt]{article}
\usepackage{jheppub} % for details on the use of the package, please                    % see the JHEP-author-manual
\usepackage{amsfonts,amsmath,amssymb,epsf}
\usepackage{amsmath,braket}
\usepackage{graphicx,hyperref,color}
\hypersetup{
    bookmarks=true,         % show bookmarks bar?
    unicode=false,          % non-Latin characters in Acrobat’s bookmarks
    pdftoolbar=true,        % show Acrobat’s toolbar?
    pdfmenubar=true,        % show Acrobat’s menu?
    linktocpage=true,       % hyperlink by page number in contents
    pdffitwindow=false,     % window fit to page when opened
    pdfstartview={FitH},    % fits the width of the page to the window
    pdfnewwindow=true,      % links in new PDF window
    colorlinks=true,       % false: boxed links; true: colored links
    linkcolor=blue,          % color of internal links (change box color with linkbordercolor)
    citecolor=blue,        % color of links to bibliography
    filecolor=blue,      % color of file links
    urlcolor=blue           % color of external links
}

\numberwithin{equation}{section}									% equation numbering by section

\newcommand{\de}{\partial}
\newcommand{\be}{\begin{equation}}
\newcommand{\ba}{\begin{eqnarray}}
\newcommand{\ea}{\end{eqnarray}}
\newcommand{\ee}{\end{equation}}

\newcommand{\s}{\sqrt}

\newcommand{\ti}{\tilde}
\newcommand{\ap}{\alpha}

\newcommand{\ddd}{\cdot\cdot\cdot}
\newcommand{\no}{\nonumber \\}

\newcommand{\bea}{\begin{eqnarray}}
\newcommand{\eea}{\end{eqnarray}}
\newcommand{\bes}{\begin{equation*}}
\newcommand{\beas}{\begin{eqnarray*}}
\newcommand{\eeas}{\end{eqnarray*}}
\newcommand{\bas}{\begin{array*}}
\newcommand{\eas}{\end{array*}}
\newcommand{\ees}{\end{equation*}}

\newcommand{\ep}{\epsilon}

\newcommand{\eg}{{\it e.g.,}\ }
\newcommand{\ie}{{\it i.e.,}\ }
\newcommand{\viz}{{\it viz,}\ }
\newcommand{\mt}[1]{\textrm{\tiny #1}}
\renewcommand{\(}{\left(}
\renewcommand{\)}{\right)}
\renewcommand{\[}{\left[}
\renewcommand{\]}{\right]}

\newcommand{\GN}{G_\mt{N}}

\newcommand{\tu}{\tilde{u}}
\newcommand{\tv}{\tilde{v}}
\newcommand{\tz}{\tilde{z}}
\newcommand{\tT}{\tilde{T}}
\subheader{\today}
\title{\boldmath Gluing AdS/CFT}

\author[a]{Taishi Kawamoto,}
\author[a]{Shan-Ming Ruan,}
\author[a,b,c]{Tadashi Takayanagi}

\affiliation[a]{Center for Gravitational Physics and Quantum Information, Yukawa Institute for Theoretical Physics, Kyoto University,\\
	Kitashirakawa Oiwakecho, Sakyo-ku, Kyoto 606-8502, Japan}
\affiliation[b]{Inamori Research Institute for Science,\\
620 Suiginya-cho, Shimogyo-ku,Kyoto 600-8411 Japan}

\affiliation[c]{Kavli Institute for the Physics and Mathematics
of the Universe (WPI),\\
University of Tokyo, Kashiwa, Chiba 277-8582, Japan}

\emailAdd{taishi.kawamoto@yukawa.kyoto-u.ac.jp}
\emailAdd{ruan.shanming@yukawa.kyoto-u.ac.jp}
\emailAdd{takayana@yukawa.kyoto-u.ac.jp}

\abstract{In this paper, we investigate gluing together two Anti-de Sitter (AdS) geometries along a timelike brane, which corresponds to coupling two brane field theories (BFTs) through gravitational interactions in the dual holographic perspective. By exploring the general conditions for this gluing process, we show that the energy stress tensors of the BFTs backreact on the dynamical metric in a manner reminiscent of the TTbar deformation. In particular, we present explicit solutions for the three-dimensional case with chiral excitations and further construct perturbative solutions with non-chiral excitations. 
}

\begin{document} 

%%%%%%%%%%%%%%%%%%%%%%%%%%%%%%%%%%%%%%%%%%%%%%%%%%%%
\begin{flushright}
YITP-23-27
\end{flushright}
%%%%%%%%%%%%%%%%%%%%%%%%%%%%%%%%%%%%%%%%%%%%%%%%%%%%
	\maketitle
	\flushbottom

%%%%%%%%%%%%%%%%%%%%%%%%%%%%%%%%%%%%%%%%%%%%%%%%%%%%%%%%
%%%%%%%%%%%%%%%%%%%%%%%%%%%%%%%%%%%%%%%%%%%%%%%%%%%%%%%%
\section{Introduction}
\label{sec:intro}
%%%%%%%%%%%%%%%%%%%%%%%%%%%%%%%%%%%%%%%%%%%%%%%%%%%%%%%%
%%%%%%%%%%%%%%%%%%%%%%%%%%%%%%%%%%%%%%%%%%%%%%%%%%%%%%%%
The AdS/CFT correspondence has been a remarkable tool for understanding the properties of quantum gravity from field theories and for analyzing strongly interacting field theories through classical gravity calculations \cite{Maldacena:1997re}. This duality relates quantum gravity on $d+1$ dimensional anti-de Sitter spaces (AdS$_{d+1}$) to a class of conformal field theories (CFT$_d$) residing on the boundary of AdS$_{d+1}$ bulk spacetime. In this sense, the AdS/CFT correspondence can be viewed as a special example of holography principle \cite{tHooft:1993dmi,Susskind:1994vu}, which is a powerful and fundamental idea that quantum gravity on various spacetimes can be described by theories of quantum matter.

To gain a deeper understanding of the quantum origin of the Universe, one may be tempted to extend the AdS/CFT correspondence to more realistic spacetimes, such as de Sitter spaces. However, this is a highly non-trivial problem, mainly because such cosmological spacetimes typically lack timelike boundaries where the dual field theory could reside. Consequently, identifying the non-gravitational theory that is dual to gravity in cosmological spacetime becomes exceedingly difficult. Several approaches have been taken to address this conundrum. In the case of de Sitter holography, the first idea is to employ the spacelike boundaries in de Sitter spaces, which is referred to as the dS/CFT correspondence \cite{Strominger:2001pn,Witten:2001kn,Maldacena:2002vr}. Other approaches include, \eg the dS/dS duality \cite{Alishahiha:2004md,Dong:2018cuv}, the surface/state duality \cite{Miyaji:2015yva}, static patch holography \cite{Susskind:2021dfc,Susskind:2021esx}, and the von-Neumann algebras \cite{Chandrasekaran:2022cip}. Each of these approaches presents unique challenges and opportunities, and further research is hopeful to yield fascinating insights into the nature of quantum gravity and its relationship to our cosmology.

 The primary purpose of this paper is to initiate the exploration of the concept of ``holography without boundaries" through the modification of the AdS/CFT duality. In the conventional AdS/CFT correspondence, the $d+1$-dimensional bulk spacetime is dual to a conformal field theory living on its $d$-dimensional conformal boundary. Rather than proposing an entirely novel holographic duality, we modify the original AdS/CFT framework by gluing two distinct portions of AdS geometries, which are enclosed by the timelike boundaries $\Sigma^{(1)}$ and $\Sigma^{(2)}$, respectively. Subsequently, we join the two AdS$_{d+1}$ spacetimes together along the timelike hypersurface by identifying the two branes, \ie $\Sigma^{(1)}=\Sigma^{(2)}(\equiv\Sigma)$, which could create an AdS bulk spacetime without boundaries. We anticipate that the resulting bulk geometry will be dual to two lower-dimensional field theories interacting through induced dynamical gravity on the braneworld $\Sigma$. In this work, we provide a detailed description of gluing AdS/CFT, with a particular focus on the AdS$_3/$CFT$_2$ case.

This framework bears some resemblance to the brane-world models \cite{Randall:1999ee,Randall:1999vf,Gubser:1999vj,Karch:2000ct}, which assert that a $d+1$ dimensional AdS geometry with a finite cut-off is dual to a conformal field theory coupled to a certain quantum gravity on the $d$ dimensional boundary of the AdS$_{d+1}$. In such models, one imposes the Neumann boundary condition on the boundary surface $\Sigma$, which is referred to as the end-of-the-world brane. When the boundary is an AdS$_d$, this is also interpreted as the gravity dual of a CFT on a manifold with boundaries, so called the AdS/BCFT correspondence \cite{Karch:2000gx,Takayanagi:2011zk,Fujita:2011fp}. 
It is notable that our trivial class of gluing AdS solutions with vanishing stress tensors can be reduced to two copies of the AdS geometry with the end-of-the-world brane. Gluing two AdS/BCFT geometries partially along a common AdS boundary $\Sigma$ has been studied by many authors in the context of the gravity duals of defect or interface CFT \cite{Erdmenger:2014xya,Erdmenger:2015spo,Bachas:2020yxv,Bachas:2021fqo,Simidzija:2020ukv,May:2021xhz,Anous:2022wqh,Loran:2010qn,Pasquarella:2022ibb}, the Janus solutions \cite{Aharony:2003qf,Bak:2003jk,DHoker:2007zhm,Bak:2007jm,Azeyanagi:2007qj} and also recent developed double holography \cite{Chen:2020uac,Chen:2020hmv,Geng:2020fxl} (refer also to \cite{Martelli:2001tu} for a RG flow setup). Also, the idea of coupling an AdS to another spacetime via the AdS boundary can be found in the context of island formula associated with black hole evaporation  \cite{Penington:2019npb,Almheiri:2019psf,Almheiri:2019hni}.

It is also intriguing to note that our models of gluing two AdS/CFT 
are closely related to the wedge holography \cite{Akal:2020wfl}. The wedge holography establishes a connection between the wedge-shaped region in AdS$_{d+2}$ and quantum gravity on its boundary, which consists of two AdS$_{d+1}$ geometries. This, in turn, is dual to a $d$-dimensional CFT residing on the tip of the $d+2$-dimensional wedge, via further application of the AdS/CFT correspondence. In the middle picture of this chain of holography, two AdS$_{d+1}$ geometries are united along their boundaries, which appears similar to our gluing AdS/CFT set-up. However, the original wedge holography assumes the Dirichlet boundary condition at the $d$-dimensional tip, while in our joint spacetime, we impose the Neumann boundary condition, and hence gravity is dynamic at the tip. We proceed to examine how this gravity interacts with the energy stress tensors of the two field theories on the brane. We concentrate our detailed computations on the scenario where the end-of-the-world brane has the critical tension ($T=1$ for $d=2$). 

This paper is organized as follows: In section \ref{sec:gluing}, we present a general formulation for gluing AdS/CFT. In section \ref{sec:Cgluing}, we put forth solutions in which only chiral modes are excited. In section \ref{sec:Ngluing}, we delve into perturbative solutions in the presence of both chiral and anti-chiral excitations. In section \ref{sec:Wedge}, we explore another approach to gluing AdS/CFT by utilizing the wedge holography. Finally, in section \ref{sec:Conclusion}, we discuss potential future directions.

%%%%%%%%%%%%%%%%%%%%%%%%%%%%%%%%%%%%%%%%%%%%%%%%%%%%%%%%
%%%%%%%%%%%%%%%%%%%%%%%%%%%%%%%%%%%%%
\section{Formulation of Gluing AdS/CFT}\label{sec:gluing}
%%%%%%%%%%%%%%%%%%%%%%%%%%%%%%%%%%%%%%%%%%%%%%%%%%%%%%%%
%%%%%%%%%%%%%%%%%%%%%%%%%%%%%%%%%%%%%%%%%%%%%%%%%%%%%%%%
\begin{figure}[ht]
  \centering
   \includegraphics[width=5in]{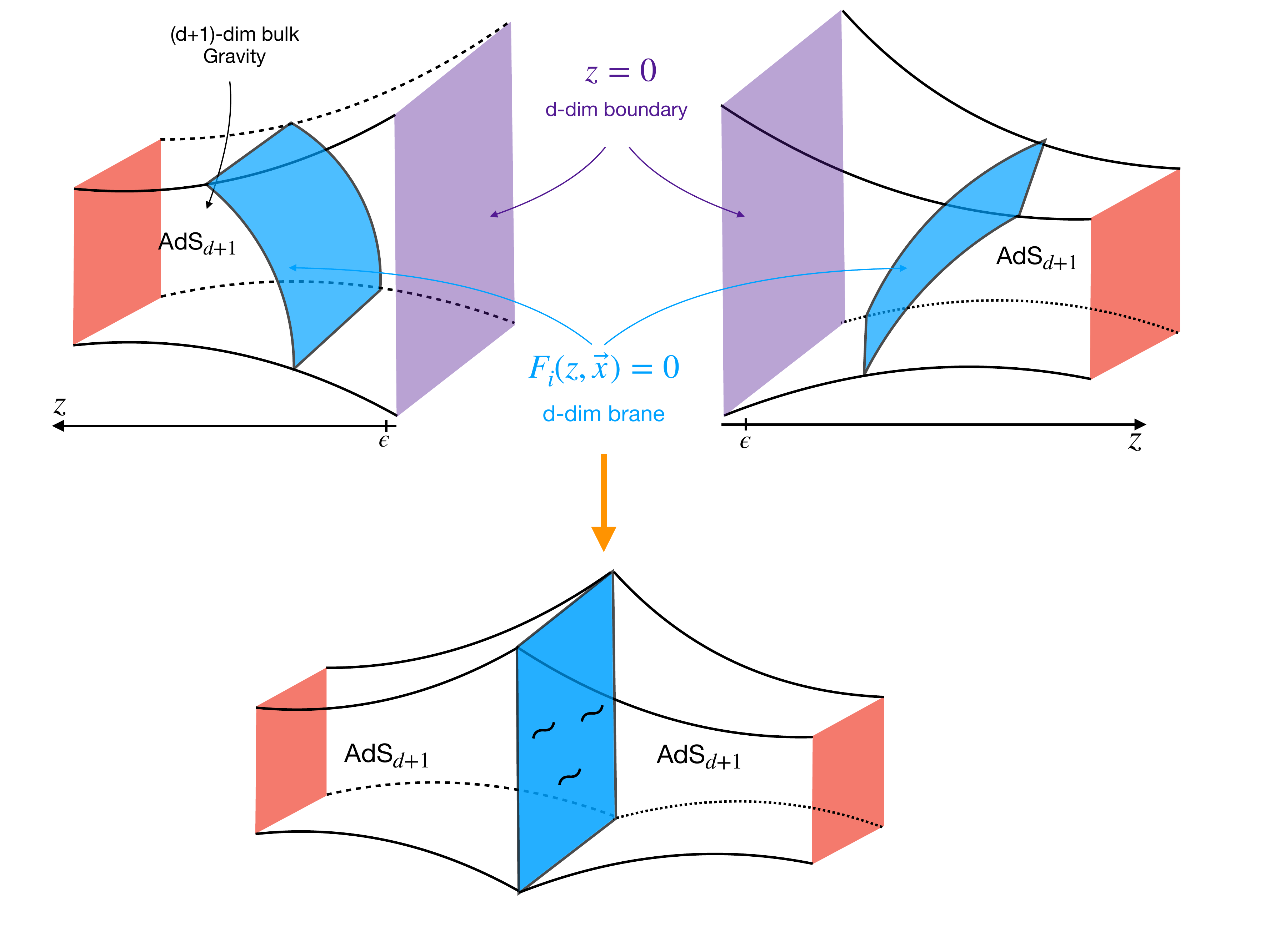}
  \caption{Our setup for gluing two AdS/CFT: We introduce a brane into each AdS bulk spacetime and remove a portion outside the brane. The new spacetime is formed by joining the remaining bulk spacetime along the timelike brane represented by the blue surface in the figure.} \label{fig:setup}
\end{figure}

In this section, we illustrate the basic constraints for gluing two AdS bulk spacetimes along a codimension-one (timelike) hypersurface that is denoted by $\Sigma$. In this paper, we assume the presence of the pure gravity in AdS$_{d+1}$ bulk spacetime. As usual, the bulk gravity theory for each side is given by standard Einstein gravity with a negative cosmological constant. Thus, the total action of the bulk spacetime is represented as follows:
\begin{equation}
I_{\text {bulk}}=\sum_{a} \frac{1}{16 \pi \GN } \int_{\mathrm{bulk }} d^{d+1} y \sqrt{-g}\left(   \mathcal{R}\left[g_{\mu \nu}\right] +\frac{d(d-1)}{L^2_{\mt{AdS}_a}}\right)\,,
\end{equation}
where $L_{\mt{AdS}_a}$ with $a=1,2$ denotes the corresponding AdS radius for two AdS bulk spacetimes, respectively. We begin by introducing a codimension-one brane to separate the two independent AdS bulk spacetimes. 
For the sake of simplicity, the brane is characterized by a fixed tension term in the main text. As a result, the boundary term in the total action thus consists of not only the standard GHY boundary terms but also a tension term, \viz 
\begin{equation}
I_{\rm{bdy}}=
\frac{1}{8 \pi \GN } \int_{\mathrm{brane}_1}d^2x \sqrt{-h^{(1)}}(K^{(1)}-T) + \frac{1}{8 \pi \GN } \int_{\mathrm{brane}_2}d^2x \sqrt{-h^{(2)}}(K^{(2)}-T) \,,
\end{equation}
where $T$ is the tension\footnote{We can choose different values of the tension for $a=1$ and $a=2$. However, only their sum is relevant in our analysis. Thus we choose them to take the same value $T$.}, $h^{(a)}_{ij}$ is the induced metric on the brane, and $K^{(a)}$ denotes the trace of the extrinsic curvature of the brane with respect to each side. Since there is no matter term in the bulk, the bulk spacetime still satisfies $\mathcal{R}_{\mu\nu} =  - \frac{d}{L^2_{\mt{AdS}_i}} g_{\mu\nu}^{(a)}$. The new bulk spacetime is then built by gluing the brane from two sides, as shown in Figure \ref{fig:setup}.

\subsection{Junction Conditions}
The junction condition on the brane is nothing but the so-called Israel junction conditions, \ie 
\begin{equation}\label{eq:Israel}
	\begin{split}
		h_{ij}^{(1)}  &= h_{ij}^{(2)}  \,,  \\
		[K_{ij}]-  [K] h_{ij} &= - 2 T h_{ij} \,, 
	\end{split}
\end{equation}
with $[K_{ij}]$ denoting the jump of $K_{ij}$ across the brane, namely
\begin{equation}
	[K_{ij}] = K_{ij}^{(1)}+  K_{ij}^{(2)} \,.
\end{equation}
Our definition of the extrinsic curvature is given by $K_{ij} = h^\mu_i h^\nu_j \nabla_\mu n_\nu$ with the normal vector $n_\mu$ outward pointing in both directions. It is important to note that the Israel junction conditions defined in eq.~\eqref{eq:Israel} presuppose that the coordinate systems of the brane from both sides are the same. Note that the Israel junction conditions result in the following coordinate-independent constraints:
\begin{equation}\label{eq:simplejunctions}
	\begin{split}
		R [ h^{(1)}   ] &=	R [ h^{(2)}  ] \,,\\
		K^{(1)} + K^{(2)} &= \frac{2d}{d-1} T\,. 
	\end{split}
\end{equation}
Since the bulk spacetime is the solution of the vacuum Einstein equations, momentum constraints also have been automatically satisfied, \viz 
\begin{equation}\label{eq:momentum}
D^i K_{ij}^{(a)} - D_j K^{(a)} =0 \,. 
\end{equation}
In the case of high-dimensional bulk spacetime, the two scalar functions do not suffice to completely solve the Israel junction conditions. However, most of the equations in eq.~\eqref{eq:Israel} for three-dimensional AdS$_3$ spacetime are redundant. For instance, it can be noticed that the first condition, which states the agreement of the Ricci scalar of the two-dimensional brane on both sides, is sufficient to ensure the match of the induced geometry.

\subsection{Constant-Mean-Curvature Slice in AdS spacetime}
One can imagine that the configuration of the hypersurface in a general bulk spacetime could be very complicated. However, we will focus on the special bulk spacetime, \ie the vacuum solutions of the Einstein equations with a negative cosmological constant. 
As we will demonstrate in the subsequent sections, the codimension-one brane consistently manifests as a hypersurface with constant mean curvature in the AdS bulk spacetime.

First of all, one can apply the Gauss equation to a timelike hypersurface as follows: 
\begin{equation}
	\mathcal{R} + 2 \mathcal{R}_{\mu\nu}n^\mu n^\nu = R -   K^{\mu\nu} K_{\mu\nu} + K^2 \,,
\end{equation}
 and immediately derive the Hamiltonian constraint, \viz 
\begin{equation}\label{eq:Hamiltonian}
	R = K^2 - K^{\mu\nu} K_{\mu\nu}  -\frac{d(d-1)}{L^2_{\mt{AdS}}} \,, 
\end{equation}
with using the fact that $(d+1)$-dimensional bulk spacetime is the vacuum solution with $\mathcal{R}_{\mu\nu} = - \frac{d}{L^2_{\mathrm{AdS}}}g_{\mu\nu}$. 
Consequently, it has been established that the intrinsic curvature $R$ of the hypersurface is entirely determined by its extrinsic curvature tensors. On the other hand, the second junction condition gives rise to the following two equalities: 
\begin{equation}\label{eq:junction condition 2}
	\begin{split}
		K_{ij}^{(1)}K^{(1)ij} - (K^{(1)})^2 + K_{ij}^{(2)}K^{(1)ij} - K^{(1)}K^{(2)} + 2 T  K^{(1)} &=0 \,,  \\
			K_{ij}^{(2)}K^{(2)ij} - (K^{(2)})^2 + K_{ij}^{(1)}K^{(2)ij} - K^{(1)}K^{(2)} + 2 T  K^{(2)} &=0 \,. 
		\end{split}
\end{equation} 
With using the Hamiltonian constraint, one can find that the difference of the above two equations leads to 
\begin{equation}
\begin{split}
 K^{(1)} - K^{(2)} &= \frac{1}{2T} \[ \(   (K^{(1)})^2-K_{ij}^{(1)}K^{(1)ij}\) -\(   (K^{(2)})^2-K_{ij}^{(2)}K^{(2)ij}\)\]\\  
 &= \frac{d(d-1)}{2T} \(\frac{1}{L^2_{\mt{AdS}_1}}   -  \frac{1}{L^2_{\mt{AdS}_2}}  \)\,, \\
 \end{split}
\end{equation}
 where the second equality follows from the identification of the Ricci scalar (\ie the first junction condition). 
 By incorporating the above observations with the second junction condition expressed in equation \eqref{eq:simplejunctions}, we immediately arrive at
 \begin{equation}\label{eq:Kconstant}
	\begin{split}
		K^{(1)} &= \frac{d\, T}{d-1}+\frac{d(d-1)}{4T} \(\frac{1}{L^2_{\mt{AdS}_1}}   -  \frac{1}{L^2_{\mt{AdS}_2}}  \)\,, \\
		K^{(2)} &= \frac{d \, T}{d-1}+\frac{d(d-1)}{4T} \(\frac{1}{L^2_{\mt{AdS}_2}}   -  \frac{1}{L^2_{\mt{AdS}_1}}  \)\,. \\
		\end{split}
	\end{equation}
As advertised before, this ultimately leads to the conclusion that the codimension-one brane on either side is always a hypersurface with a constant mean curvature. It is noteworthy that the two equations with respect to two sides of the brane are independent of each other, which is different from the original Israel junction conditions presented in equation \eqref{eq:Israel}. Additionally, if $L_{\mt{AdS}_1} = L{\mt{AdS}_2}$, a more symmetrical setup is achieved due to 
\begin{equation}\label{eq:K=2}
K^{(1)} = K^{(2)} = \frac{d \, T}{d-1} \,.
\end{equation}

\subsection{Hamiltonian Constraint and $T\bar{T}$ deformation on the brane}
Focusing on the geometry of the codimension-one brane, the variation of the total action reads 
\begin{equation}\label{eq:tauij}
    \delta I_{\rm{bulk}} + \delta I_{\rm{bdy}}  =  \frac{1}{8\pi \GN}  \int \sqrt{-h} d^dx \( K_{ij}^{(1)} - K^{(1)}h _{ij}+K_{ij}^{(2)} - K^{(2)}h _{ij} + 2Th_{ij}  \)\delta h^{ij} \,. 
\end{equation}
 With respect to the d-dimensional metric $h_{ij}$, one can also interpret the Israel junction as the Einstein equation on the brane, \ie 
\begin{equation}\label{eq:Einsteinbrane}
 \tau^{(1)}_{ij} + \tau^{(2)}_{ij} =0\,,
\end{equation}
where we have defined two distinct stress tensors on the brane in terms of 
\begin{equation}
\begin{split}
   \tau_{ij}^{(a)}:&=   K_{ij}^{(a)} - K^{(a)}h _{ij} + T^{(a)}h_{ij}  \,, \\
   \end{split}
\end{equation}
with $T^{(1)}+T^{(2)}=2T$. This definition resembles the renormalized Brown-York stress tensor (or the holographic boundary stress tensor) in the conventional AdS/CFT correspondence. For $d=2$, they are proportional to each other with a negative coefficient, as we will see below. The trace of the brane stress tensor can easily be obtained as the following:
\begin{equation}
 \tau^{(a)} \equiv  \tau_{ij}^{(a)} h^{ij} = d \, T^{(a)} - (d-1)K^{(a)}  \,. 
\end{equation}
First of all, one can notice that the brane stress tensors are conserved, \viz 
\begin{equation}
    D^i \tau_{ij}^{(a)} = D^i K_{ij}^{(a)}-D_j K^{(a)} =0  \,,
\end{equation}
thanks to the momentum constraint on the brane as shown in eq.~\eqref{eq:momentum}.
We are interested in the expectation value of $T\bar{T}$ operator with respect to the brane stress tensor $\tau_{ij}$, \ie 
\begin{equation}
    \langle \tau \bar{\tau}  \rangle \equiv  \langle  \tau^{ij} \rangle \langle \tau_{ij} \rangle - \langle \tau^i_i \rangle^2\,,
\end{equation}
for two-dimensional field theories. For a generic high-dimensional bulk spacetime, the corresponding generalization is given by 
\begin{equation}
\begin{split}
 \tau^{(a)ij} \,  \tau^{(a)}_{ij}-\frac{(\tau^{(a)})^2   }{d-1} 
&\equiv  K^{(a)ij}K^{(a)}_{ij} -(K^{(a)})^2 + 2K^{(a)}T^{(a)} - \frac{d}{d-1}(T^{(a)})^2   \\
&=  K^{(a)ij}K^{(a)}_{ij} - (K^{(a)})^2+ \frac{d}{d-1}(T^{(a)})^2 + 2 T^{(a)} \( K^{(a)}- \frac{d}{d-1} T^{(a)} \)  \,, 
\end{split}
\end{equation}
where we have recast the last term as the trace of the stress tensors. With this redefinition, we can rewrite the Hamiltonian constraint derived in eq.~\eqref{eq:Hamiltonian} as
\begin{equation}
  R- (K^{(a)})^2 + K^{(a)ij} K^{(a)}_{ij} + \frac{d(d-1)}{L^2_{\mt{AdS}}}  = R + \mu^{(a)} + \(\tau^{(a)ij} \,  \tau^{(a)}_{ij}-\frac{(\tau^{(a)})^2   }{d-1} \) + \frac{2 T^{(a)}}{d-1} \tau^{(a)}=0 \,,
\end{equation}
by identifying the constant part as a potential term, \ie 
\begin{equation}
 \mu^{(a)} = \frac{d(d-1)}{L^2_{\mt{AdS}_a}} -  \frac{d}{d-1}(T^{(a)})^2 \,.
\end{equation}
For each side, the trace equation on the brane can be interpreted as the flow equation of the stress tensor under the so-called $T\bar{T}$ deformation, \viz
\begin{equation}
 \frac{2 T^{(a)}}{d-1} \tau^{(a)} = - R  -  \mu^{(a)} - \( \tau^{(a)ij} \,  \tau^{(a)}_{ij}-\frac{(\tau^{(a)})^2   }{d-1}  \) \,.
\end{equation}
Until this point, we have allowed for arbitrary choices of the two tension terms $T^{(a)}$. However, a more natural choice is given by 
\begin{equation}\label{eq:T1T2}
\begin{split}
     T^{(1)} = T+\frac{(d-1)^2}{4T} \(\frac{1}{L^2_{\mt{AdS}_1}}   -  \frac{1}{L^2_{\mt{AdS}_2}}  \)\,, \\
     T^{(2)} = T+\frac{(d-1)^2}{4T} \(\frac{1}{L^2_{\mt{AdS}_2}}   -  \frac{1}{L^2_{\mt{AdS}_1}}  \)\,.
\end{split}
\end{equation}
As a result, the second junction condition implies the traceless condition of the boundary stress tensor, namely 
\begin{equation}
 \tau^{(a)} = d \, T^{(a)} - (d-1)K^{(a)} =0 \,. 
\end{equation}
We would like to note that this traceless condition is realized regardless of the particular choice of the value of tension $T$. Furthermore, the potential terms are also identical, \ie  
\begin{equation}
 \mu = \mu^{(1)} =\mu^{(2)} = \frac{d(d-1)}{L^2_{\mt{AdS}_a}} -  \frac{d}{d-1}(T^{(a)})^2 \,,
\end{equation}
after taking eq.\eqref{eq:T1T2} for $T^{(a)}$. The flow equations of the two brane stress tensors, \ie the brane constraint equations, reduce to 
\begin{equation}\label{eq:constraint}
 R  +   \mu = - \( \tau^{(1)ij} \,  \tau^{(1)}_{ij}-\frac{(\tau^{(1)})^2   }{d-1} \)=- \( \tau^{(2)ij} \,  \tau^{(2)}_{ij}-\frac{(\tau^{(2)})^2   }{d-1}\) \,.
\end{equation}

\subsection{AdS$_3$ bulk spacetime}
In the remainder of the paper, we will concentrate on the case with the identical AdS radius: $L_{\mt{AdS}_1}=L=L_{\mt{AdS}_2}$ for simplicity. We will specifically focus on AdS$_3$ for constructing explicit configurations of two-dimensional branes. The geometric constraint for a timelike brane in AdS$_3$ is expressed as
\begin{equation}\label{eq:constrainAdS3}
2 T \, \tau^{(a)}= - R -   \mu  - \( \tau^{(a)ij} \,  \tau^{(a)}_{ij}- (\tau^{(a)})^2 \)  \,, 
\end{equation}
with $\mu = \frac{2}{L^2}-2T^2$. 
Let us first examine the special case where the brane is pushed to the conformal boundary before delving into the discussion of brane in the center of AdS bulk spacetime. Near the conformal boundary, we can describe the asymptotic geometry in the Fefferman–Graham gauge as follows:
\begin{equation}
    d s^2=g_{\mu \nu} d x^\mu d x^\nu= \frac{L^2}{ r^2}dr^2+\frac{r^2}{L^2}\gamma_{ij}d x^i d x^j,
\end{equation}
where the conformal boundary is located at $r \to \infty$. It is worth noting that the brane tension term with fixing $T= \frac{1}{L}$ for each side, serves as the counterterm, \viz
\begin{equation}
I_{\text {ct}}= -\frac{1}{8\pi \GN L } \int_{\rm{bdy}} d^{2} x \sqrt{-h}\,,
\end{equation}
which is used for holographic renormalization in AdS$_3$. Furthermore, the brane stress tensor $\tau^{(a)}_{ij}$ defined in eq.~\eqref{eq:tauij} thus reduces to the renormalized quasi-local stress tensor \cite{Balasubramanian:1999re}, \ie 
\begin{equation}
 \mathcal{T}^{ij}  \equiv \frac{2}{\sqrt{h}}\frac{\delta S_{\rm{ren}}}{\delta h_{ij}}=  -\frac{1}{8\pi \GN}\( K^{ij} - K h^{ij} + h^{ij}  \)  = -\frac{\tau^{\ij}}{8\pi \GN} \,,
\end{equation}
which can be interpreted as the expectation value of the stress tensor of CFT at the conformal boundary of asymptotically anti-de Sitter spacetime. For a finite cut-off surface located at $r=r_c$, we can read the boundary metric $\gamma_{ij}$ associated with the field theory from the induced metric by 
\begin{equation}
  h_{ij} \big|_{r=r_c}=  \frac{r_c^2}{L^2} \gamma_{ij}  \,.
\end{equation}
The holographic stress tensor of the boundary CFT is identical to the renormalized quasilocal stress tensor, \ie $T_{ij}=\mathcal{T}_{ij}$, for two-dimensional CFT. In the conformal limit as $r_c \to \infty$, the Hamiltonian constraint in eq.~\eqref{eq:Hamiltonian} associated with the conformal boundary agrees with the trace anomaly of two-dimensional CFT \cite{Henningson:1998gx,Henningson:1998ey}, namely
\begin{equation}
   \lim_{r_c \to 0 } \langle T^i_{i} \rangle = \frac{r_c^2}{L^2} h^{ij}\mathcal{T}_{ij}  = + \frac{r_c^2}{16 \pi \GN L } R[h_{ij}] = + \frac{c}{24\pi}R[\gamma_{ij}]\,, 
\end{equation}
after taking $c = \frac{3L}{2 \GN}$. This is a typical scenario in the AdS$_3$/CFT$_2$ correspondence, where we enforce the Dirichlet boundary condition $\delta \gamma_{ij}=0$ on the conformal boundary. Considering a finite cut-off surface, the corresponding field theory is deformed by the $T\bar{T}$ term \cite{Zamolodchikov:2004ce,Smirnov:2016lqw,Cavaglia:2016oda,Cardy:2018sdv}. As a result, the Hamiltonian constraint \eqref{eq:constrainAdS3} becomes the $T\bar{T}$ flow equation \cite{McGough:2016lol,Hartman:2018tkw}, \ie 
\begin{equation}\label{eq:flow equation}
  		\langle T^i_{i} \rangle
  		= \frac{c}{24\pi}\, R[\gamma] + \frac{\lambda}{4} \big(\langle T^{ij}\rangle \langle T_{ij}\rangle -\langle T^i_{i}\rangle^2\big),\\
\end{equation}
where the coupling constant $\lambda$ for $T\bar{T}$ deformation is identified as the bulk quantity via
\begin{equation}
\lambda = \frac{ 16 \pi \GN L}{  r^2_c} \,,
\end{equation}
and potential term $\mu$ vanishes since the counterterm corresponds to $T=\frac{1}{L}$. Note that $T_{ij}$ is the stress tensor associated with the deformed theory rather than CFT$_2$ on the conformal boundary. It is obvious that the $T\bar{T}$ term in terms of boundary quantities would not contribute in the limit $r_c \to \infty$ due to the appearance of the double traces.

Instead of imposing the Dirichlet boundary condition on the brane, we aim to connect two AdS$_3$ bulk spacetimes via the dynamical brane. Using the Israel junction conditions, the brane is fixed as a CMC hypersurface with $K^{(a)}= 2T$ with respect to the bulk spacetime of each side, as previously demonstrated. The possible configurations for a generic brane with tension $T$ residing in the bulk spacetime are restricted by the following equation:
\begin{equation}\label{eq:constrainAdS302}
 - 2T \tau^{(a)} = 0=R   + \mu +  \( \tau^{(a)ij} \,  \tau^{(a)}_{ij}- (\tau^{(a)})^2    \)   =R  +  \mu+   \tau^{(a)ij}\tau^{(a)}_{ij}   \,,
\end{equation}
where the trace of the brane stress tensors vanishes due to the CMC condition. It differs from the normal story of $T\bar{T}$ deformed CFT on a finite cut-off surface.

First of all, let us think about stretching the brane to the conformal boundary. We note that in AdS$_3$, the trace of the extrinsic curvature of the conformal boundary is always fixed as 
\begin{equation}
     K^{(a)} \big|_{\rm{bdy}}= \frac{2}{L} \,.
\end{equation}
As a result, the CMC condition, \ie the junction condition, would be satisfied if and only if $T= \frac{1}{L}$. It is straightforward to see that the potential term $\mu$ vanishes after taking $T=\frac{1}{L}$, which leads us to the constraint equation: 
\begin{equation}\label{eq:constrainAdS3}
R  + \( \tau^{(a)ij} \,  \tau^{(a)}_{ij}- (\tau^{(a)})^2    \)   =R  +    \tau^{(a)ij}\tau^{(a)}_{ij}  =0 \,.
\end{equation}
In other words, it can be ascertained that the aforementioned constraint prevents us from gluing any two arbitrary AdS$_3$ bulk spacetimes along a timelike brane. In the next section, we will proceed to find solutions for the profiles of the brane by explicitly solving the junction conditions. Given that the radii of the two AdS bulk spacetimes have been selected to be congruent, we will set $L=1$ throughout the remainder of the present paper.

%%%%%%%%%%%%%%%%%%%%%%%%%%%%%%%%%%%%%%%%%%%%%%%%%%%%%%%%
%%%%%%%%%%%%%%%%%%%%%%%%%%%%%%%%%%%%%
\section{Gluing AdS$_3/$CFT$_2$}
 \label{sec:Cgluing}
%%%%%%%%%%%%%%%%%%%%%%%%%%%%%%%%%%%%%%%%%%%%%%%%%%%%%%%%

\begin{figure}[t]
  \centering
   \includegraphics[width=10cm]{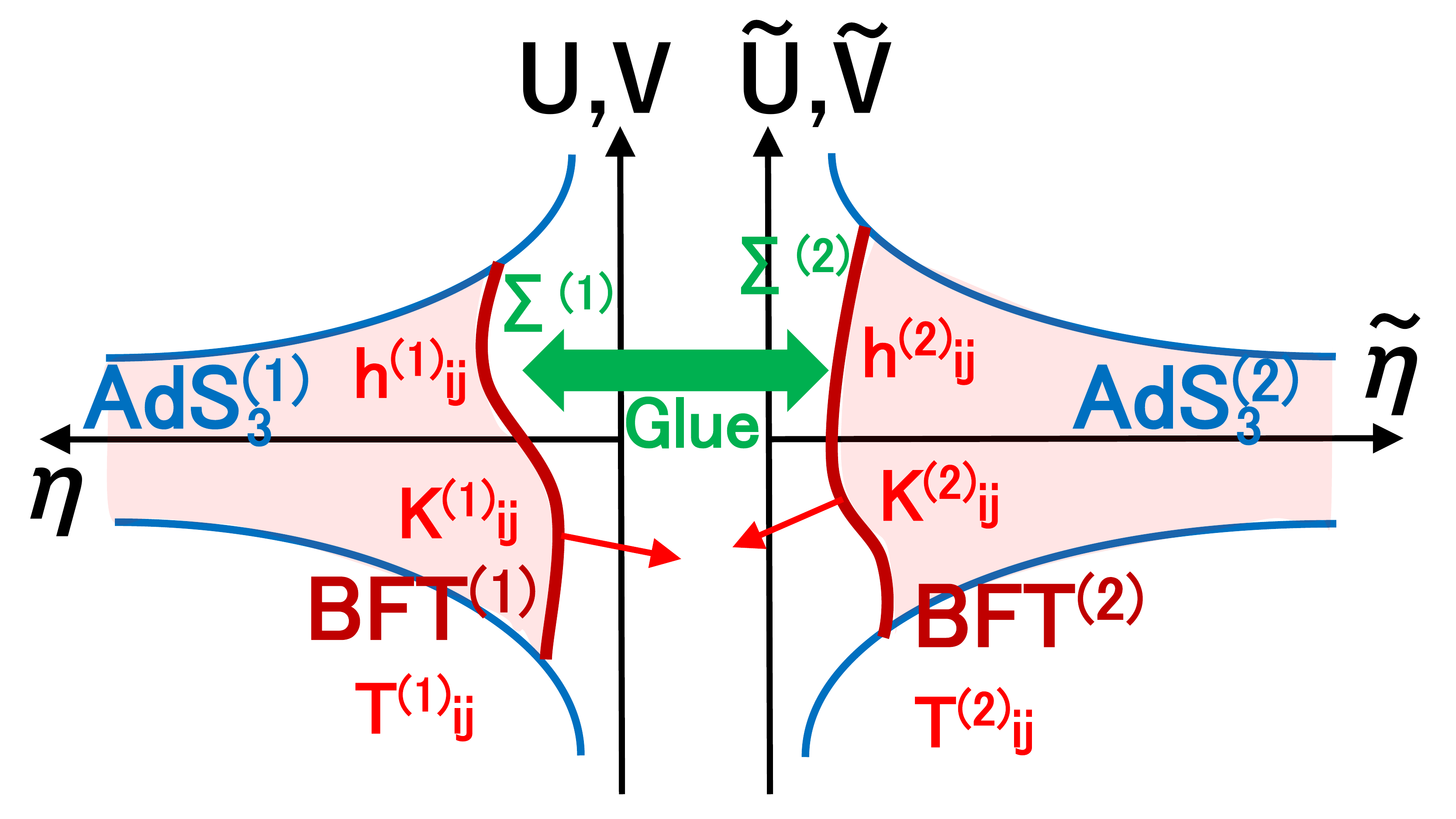}
  \caption{Gluing two Poincar\'e AdS geometries along timelike hypersurfaces 
  $\Sigma^{(1)}$ and $\Sigma^{(2)}$ and our notations for various physical quantities.} 
\label{fig:glueAdS3}
\end{figure}

%%%%%%%%%%%%%%%%%%%%%%%%%%%%%%%%%%%%%%
To construct explicit solutions of joint AdS background, we mainly focus on solutions obtained by gluing two AdS$_3$ geometries together. In this section, we analyze an exactly solvable class of solutions with chiral excitations. We denote the boundary of the left-sided and that of the right-sided AdS before gluing, as $\Sigma^{(1)}$ and $\Sigma^{(2)}$, respectively. Correspondingly, the effective field theory living on each brane is presented by BFT$^{(1)}$, BFT$^{(2)}$ as an abbreviation of the brane field theory. 
As depicted in Figure \ref{fig:glueAdS3}, we joint the two bulk spacetimes by gluing the two branes, which couples BFT$^{(1)}$ with BFT$^{(2)}$.
%%%%%%%%%%%%%%%%%%%%%%%%%%%%%%%%%%%%%%%%%%%%%%%%%%%%%%%%

\subsection{Symmetric Solutions}\label{sec:symmetric}
Although it is not easy to get the most general solutions of 
the Israel junction condition, the junction condition reduces to the simplest case, \ie 
\begin{equation}
\tau_{ij}^{(a)}=  K_{ij}^{(a)} - K^{(a)}h _{ij} + T h_{ij} =0 \,, \\
\end{equation}
or equivalently 
\begin{equation}\label{eq:simple}
  K_{ij}^{(a)} = \frac{Th_{ij}}{d-1} \,,
\end{equation}
when the left and right regions are exactly symmetric. It is a stronger constraint of our first conclusion that the brane is a CMC slice with $K^{(a)}=\frac{Td}{d-1}$ for the left/right bulk spacetime. It is obvious that the junction condition \eqref{eq:Einsteinbrane} is thus the same as the Neumann boundary condition for each side, which is explicitly used for the construction of AdS/BCFT correspondence \cite{Takayanagi:2011zk}. Supposing the bulk spacetime is given by the vacuum solution of Einstein equation with a negative constant, one can substitute $\mathcal{R}_{\mu\nu} =  -dg_{\mu\nu}$ to the contracted Gauss equation \eqref{eq:Hamiltonian} (Hamiltonian constraint) and immediately obtain the Ricci scalar of the d-dimensional brane, \ie 
\begin{equation}
R=\frac{d\,T^2}{d-1}  -d(d-1) \equiv -\mu\,.
\end{equation}
This is, of course, the constraint equation \eqref{eq:constraint} but with a vanishing $T\bar{T}$ term. We can regard this symmetric class of solutions as the vacuum ones because the holographic stress tensor vanishes. From this, we can conclude that the sign of the cosmological constant of the braneworld is also determined by the tension of the brane in this symmetric case. For example, the flat brane is obtained when the tension is given by the critical case with $T= d-1$. On the contrary, the AdS brane can exist with a lower tension $ |T| \le d-1$. 
Moreover, for $|T|\geq d-1$, we find the brane takes the form of a de Sitter space.

It is noteworthy that one can exactly solve equation \eqref{eq:simple} in pure AdS bulk spacetimes. As a warm-up, we begin by considering AdS$_3$ in Poincar\'e coordinates, namely 
	\begin{equation}
		ds^2 = \frac{- dt^2 + d x^2 + d \eta^2 }{\eta^2}  \,. 
  \label{pointh}
	\end{equation}
The codimension-one hypersurface in AdS$_3$ is thus parameterized by a scalar function $F(t, x, \eta)=0$. After some algebras, one can find that the hypersurface satisfying $K_{ij} \propto h_{ij}$ is solved by  \begin{equation}\label{eq:CMCAdS3}
		F(t, x, \eta) = A \(  x^2 +\eta^2 -t^2  \) + B  \eta + C  x + D t +E=0 \,,
	\end{equation}
with $(A, B, C, D, E)$ as real constants. However, we need to note that this family of solutions only depends on four free parameters, \eg $(A/E, B/E, C/E, D/E)$. In this paper, we are more interested in timelike hypersurfaces, which should satisfy the following constraint:
\begin{equation}
	\text{Timelike}: \quad 	n^\mu n_\mu = +1 >0 \,, \qquad \longrightarrow  \qquad  B^2 + C^2 -D^2 -4AE  > 0 \,.
\end{equation}
One can also work out the general solutions of eq.~\eqref{eq:simple} in other AdS$_3$ spacetime by taking the solutions shown in \eqref{eq:CMCAdS3} and performing the corresponding coordinate transformations.   
As we advertised before, one can easily check that the extrinsic curvature of the hypersurface parametrized by $F(t,x,\eta)=0$ satisfies eq.~\eqref{eq:simple}. More explicitly, we have  
\begin{equation}
	K_{ij} = \frac{\pm B  \, h_{ij} }{\sqrt{|B^2 + C^2 -D^2 -4AE  |}} \,, 
\end{equation}
where the sign depends on our choice of physical region. Obviously, it is nothing but the solution of the symmetric junction condition after taking 
\begin{equation}
T= \frac{\pm B }{\sqrt{|B^2 + C^2 -D^2 -4AE |}} \,. 
\end{equation}
In particular, we stress that the induced geometry of the hypersurface is still maximally symmetric, \ie AdS$_2$, dS$_2$ or Minkowski spacetime. One can check that the Ricci scalar of the induced metric reads 
\begin{equation}
R = \frac{2\(4AE -C^2 +D^2\)}{B^2 + C^2 -D^2 -4AE  }  \,. 
\end{equation}
We can glue a pair of identical solutions constructed explicitly in this way. 

\begin{figure}[hhh]
  \centering
   \includegraphics[width=10cm]{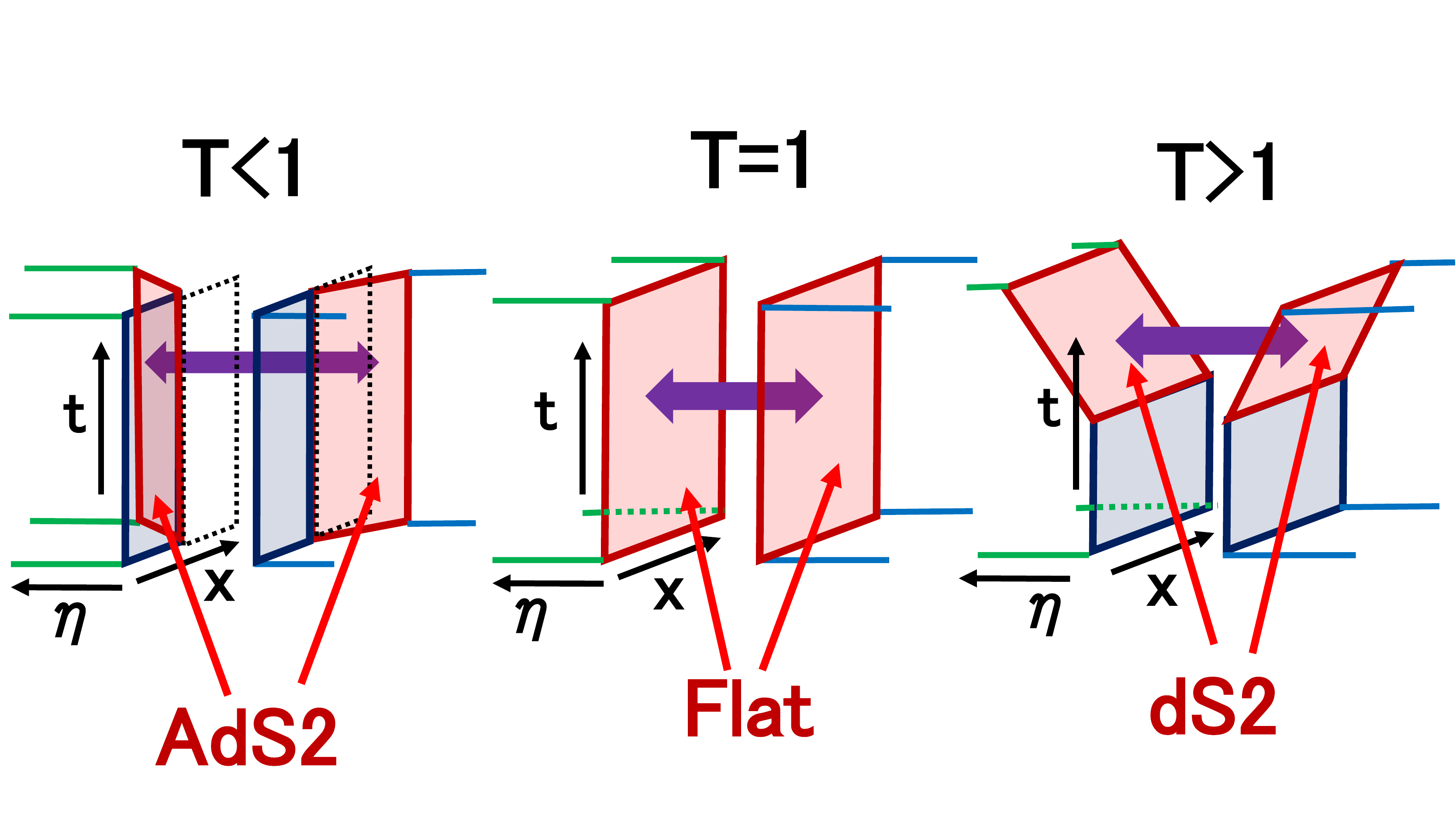}
  \caption{Sketches of gluing two AdS$_3$ geometries along 
AdS$_2$ (left), the flat space (middle), and dS$_2$ (right). The two red surfaces are glued in each case.} 
\label{fig:gluepoin}
\end{figure}

For example, the finite cut-off surface located at 
\begin{equation}
  \eta = \eta_0=\text{constant}\,,
\end{equation}
corresponds to a flat brane with tension at $T=1$. In other words, it implies that one can glue two AdS$_3$ in Poincar\'e coordinates along their finite cut-off surfaces at $\eta=\eta_0$ and $\tilde{\eta}=\tilde{\eta}_0$ by imposing the tension of the brane as $T=1$. We can find that the static timelike surface defined by $\eta=\lambda x$ is given by AdS$_2$ with $R=-\frac{2}{1+\lambda^2}$ and $T=\frac{\lambda}{\s{1+\lambda^2}}$. On the other hand, the spacelike surface with the translation invariance defined by $t=\lambda\eta$ describes dS$_2$ with $R=\frac{2}{\lambda^2-1}$ and $T=\frac{\lambda}{\s{\lambda^2-1}}$, where we assume $|\lambda|\leq 1$. We sketched the gluing of two copies of these solutions in Figure \ref{fig:gluepoin}.

\subsection{Chiral Solutions from Poincar\'e AdS$_3$}
We have shown that the brane profiles in the symmetric bulk spacetime are parametrized by eq.~\eqref{eq:CMCAdS3} thanks to the vanishing of the brane stress tensors $\tau^{(a)}_{ij}=0$. 
Different from the symmetric set-up, the two bulk spacetimes glued togehter by the brane may not be the same in general. In other words, one can expect that there are more nontrivial solutions of the brane profiles with $\tau^{(a)}_{ij} \ne 0$. Instead of directly solving the most general junction conditions \eqref{eq:Einsteinbrane}, we begin with the generalization of the previous results by including nonzero brane stress tensors but keeping a vanishing $T\bar{T}$ term, \ie $\tau^{(a)ij}\tau_{ij}^{(a)}-(\tau^{(a)})^2=0$. Correspondingly, the brane constraint equation \eqref{eq:constrainAdS3} in AdS$_3$ reduces to 
\begin{equation}\label{eq:branechiral}
  R  +   \mu =  R + 2 - 2 T^2  \,,
\end{equation}
which is the same as that in the symmetric setups. 
 For more explicit solutions, we start from Poincar\'e AdS$_3$ and denote the two bulk spacetimes as 
\begin{equation}\label{eq:Poincare02}
  g^{(1)}_{\mu\nu}dx^\mu dx^\nu= \frac{d\eta^2-dU dV}{\eta^2},\qquad g^{(2)}_{\mu\nu}dx^\mu dx^\nu=\frac{d\ti{\eta}^2-d\ti{U} d\ti{V}}{\ti{\eta}^2} \,, 
\end{equation}
where we have chosen null coordinates $(U,V), (\tilde{U}, \tilde{V})$ for later convenience. The sketch of this setup and our conventions are summarized in Figure \ref{fig:glueAdS3}. Before gluing the two AdS$_3$ bulk spacetimes, we consider two branes $\Sigma^{(a)}$ on each side by assuming the brane profiles are given by 
\begin{equation}\label{eq:chiralbrane}
    \eta=e^{-\phi(U)} \,, \qquad \ti{\eta}=e^{-\ti{\phi}(\ti{U})} \,,
\end{equation}
respectively. The induced metric of the brane $\Sigma^{(a)}$ thus reads 
\begin{equation}
   ds^2 \big|_{\Sigma^{(1)}}=  (\phi')^2dU^2 - e^{2\phi} dUdV\,,\quad  ds^2 \big|_{\Sigma^{(2)}}=  (\tilde{\phi}')^2d\tilde{U}^2 - e^{2\phi} d\tilde{U}d\tilde{V}\,,
\end{equation}
which is a two-dimensional Minkowski spacetime with $R_{ij}=0$. Note that these two coordinates on $\Sigma^{(1)}$ and $\Sigma^{(2)}$ are not simply identical in general. On the other hand, as a hypersurface of AdS$_3$, the extrinsic curvature of the brane $\Sigma^{(1)}$ is derived as 
\begin{equation}
 K_{UU} = 2\phi'^2 -\phi'' \,,\qquad   K_{VV} = 0\,,\qquad   K_{UV}= -\frac{1}{2}e^{2\phi} \,,
\end{equation}
whose trace reduces to a constant $K=2$. We have similar expressions for the second brane $\Sigma^{(2)}$. Since we have shown that the brane jointing two bulk spacetimes has to be a CMC slice, we can immediately conclude that the only possibility for gluing the two branes parametrized by the chiral form in eq.~\eqref{eq:chiralbrane} is choosing $T=1$. It is also obvious that the brane solutions shown in eq.~\eqref{eq:chiralbrane} are not the symmetric cases described in the previous subsection due to the existence of the non-vanishing brane stress tensors, \ie 
\begin{equation}
\tau^{(1)}_{UU} = -\phi''+\phi'^2 \,, \qquad \tau^{(2)}_{\tilde{U}\tilde{U}} = -\tilde{\phi}''+\tilde{\phi}'^2\,.
\end{equation}
However, this type of flux cannot curve the brane spacetime due to 
\begin{equation}
\tau^{(a)ij}\tau_{ij}^{(a)}-(\tau^{(a)})^2=0\,.
\end{equation}
As shown by the brane constraint equation \eqref{eq:branechiral}, the brane with a tension $T=1$ in this situation always leads us to a flat braneworld.

%%%%%%%%%%%%%%%%%%%%%%%%%%%%%%%%%%%%%%%%
From the above analysis, we have seen that the junction condition can be solved by taking the brane profiles as eq.~\eqref{eq:chiralbrane} and $T=1$. However, the gluing of the two flat branes $\Sigma^{(a)}$ is still nontrivial since we need to carefully match the two coordinate systems. First of all, we assume that the transformations are given by
\begin{equation}\label{eq:ansatz}
    \ti{U}=P(U),\quad \ti{V} = V + Q(U) \,, 
\end{equation}
where $P(U),Q(U)$ are functions depending on only $U$.
With this ansatz, we focus on analytically solving the Israel junction conditions in the following. The identification of the two induced metrics $h_{ij}^{(1)}=h_{ij}^{(2)}$ yields\footnote{One may notice that the first equation can only hold when $P'(U)>0$. However, we can absorb the sign or any constant by changing the ansatz in eq.~\eqref{eq:ansatz} to $\ti{V} = c_1 V + Q(U)$.}
\begin{equation}\label{eq:chiral01}
    \begin{split}
      e^{2\phi(U)} &= e^{2\ti{\phi}(P(U))}P'(U) \,, \\
    \left(\frac{d\phi(U)}{d U}\right)^2 &= \left(\frac{d \ti{\phi}(\ti{U})}{d\ti{U}}\right)^2 (P'(U))^2 -e^{2\ti{\phi}(P(U))} Q'(U)P'(U)\,.
    \end{split}
\end{equation}
In the following, we choose to work on $(U,V,\eta)$ coordinates. By substituting the first equation with the second, we obtain
\begin{equation}\label{eq:Qprime}
    Q'(U) = e^{-2\phi}\left(-\frac{P''}{P'}\frac{d\phi}{dU} + \frac{1}{4} \frac{P''^2}{P'^2}\right) \,,
\end{equation}
which relates the two functions $P(U)$ and $Q(U)$. On the other hand, we can find that the brane stress tensors $\tau_{ij}^{(2)}$ on the brane in terms of $(U,V)$ coordinates are recast as 
\begin{equation}
 \begin{split}
     \tau_{UU}^{(2)}  &= \left(-\frac{d^2}{d^2 \ti{U}}\ti{\phi}+\left( \frac{d}{d\ti{U}}\ti{\phi}\right)^2\right)P'(U)^2= -\phi''+\phi'^2-T_+(U) \,,\\
     \tau_{UV}^{(2)}&=0=\tau_{VV}^{(2)} \,,\\
 \end{split}
\end{equation}
with 
\begin{equation}\label{eq:tau12}
 T_+(U) =-\frac{1}{2}\{P,U\} = \frac{3}{4}\left(\frac{P''(U)}{P'(U)}\right)^2 -\frac{P'''(U)}{2P'(U)}\,.
\end{equation}
The second junction condition $\tau_{ij}^{(1)}+\tau_{ij}^{(2)}=0$ then yields  
\begin{equation}
 -\phi''+\phi'^2= \frac{1}{2} T_+(U) = -\frac{1}{4} \left\{\tilde{U}, U\right\}\,,\label{eomphg}
\end{equation}
which indicates that the coordinate transformation $\tilde{U}=P(U)$ is fixed by the choice of the brane profile, \ie the chiral function $\phi(U)$. 
After gluing the two branes $\Sigma^{(a)}$ with the Israel junction conditions, the non-vanishing stress tensors reduce to
\begin{equation}
\begin{split}
  \tau^{(1)}_{UU}  &= \frac{1}{2}T_+(U) = -\tau^{(2)}_{UU}   \,, \\ 
  \tau^{(2)}_{\ti{U}\ti{U}}  &= -\frac{1}{2P'^2}T_+(U)\,.
\end{split}
\end{equation}
It is worth noting that these energy stress tensors $\tau^{(a)}_{ij}$ are not the ones that are generated by a conformal transformation in a conventional way due to the extra factor $1/2$. In the following, we analyze several simple examples with the goal of deriving the explicit solutions of the brane profiles.

\subsubsection*{Vanishing Stress Tensor}
We commence our analysis with the case in which the stress tensor vanishes, i.e., $\tau_{ij}^{(1)}=0$, corresponding to the symmetric configuration discussed in the previous subsection. We first note that the equation of motion \eqref{eomphg} can be recast as 
\begin{equation}
\frac{d^2}{d^2 U}e^{-\phi}=\frac{1}{2}T_+(U)e^{-\phi}\,.
\end{equation}
With taking $\tau_{ij}=0=T_+(U)$, we can easily obtain the solutions for the brane profile $\Sigma^{(1)}$ by 
\begin{equation}
\eta = e^{-\phi} = C_1 U+C_2\,,
\end{equation}
where $C_1$ and $C_2$ are arbitrary constants. It is apparent that this type of solution coincides with those derived in equation \eqref{eq:CMCAdS3} upon assuming $\eta = e^{- \phi (u)}$. Due to the vanishing of the Schwarzian derivative defined in eq.~\eqref{eq:tau12} associated with $\tau_{UU}^{(1)}$, the coordinate transformation $P(U)$ between $\Sigma^{(1)}$ and $\Sigma^{(2)}$ is fixed to be
\begin{equation}\label{eq:tildeU}
\tilde{U}= P(U) = \frac{a U+ b}{c U+d} ,,\quad \text{with}\quad ad-bc=1 \,.
\end{equation}
This transformation corresponds to an $SL(2,\mathbf{R})$ transformation, namely, half of the isometries of the AdS$_3$ bulk spacetime. However, it is worth noting that $\tilde{U}$ defined by equation \eqref{eq:tildeU} does not cover all real values, indicating that we are gluing a portion of $\Sigma^{(2)}$ to the entire brane $\Sigma^{(1)}$. This can be traced back to the asymptotic symmetry breaking of global isometries of AdS$_3$ induced by the existence of the brane located at a finite radius. Nonetheless, 
there are still isometries left, \ie $P(U)=a U+b \in (-\infty,
 +\infty)$, under which the two branes $\Sigma^{(1)}$ and $\Sigma^{(2)}$ are equivalent. For instance, the brane profile of $\Sigma^{(2)}$ can be derived as
\begin{equation}
\begin{split}
\tilde{\phi} &= \phi - \frac{1}{2}\log a \,, \\
\tilde{\eta}&= e^{-\tilde{\phi}} = \sqrt{a} \eta \,,
\end{split}
\end{equation}
which can be understood as the profile of $\Sigma^{(1)}$ under an isometric transformation.

\subsubsection*{Constant Energy Flux} 
Furthermore, let us consider the case with a constant energy flux in the first conformal field theory, \ie 
\begin{equation}
\tau^{(1)}_{UU}=-\frac{1}{4}\{\ti{U},U\}=\ap^2\,.  
\end{equation}
This choice can be realized by selecting the function $P(U)$ as follows 
\begin{equation}
\ti{U}=P(U)=e^{2\s{2}\ap U}\,. 
\end{equation}
Solving the differential equation \eqref{eomphg} yields 
\begin{equation}
\phi(U)=-\log \left(C_1 e^{\ap U}+ C_2 e^{-\ap U}\right)\,.
\end{equation}
Notably, we have $\ti{U}=P(U)>0$ on the brane $\Sigma^{(2)}$. As a consequence, we can only glue a portion of $\Sigma^{(2)}$ with $\Sigma^{(1)}$ while keeping the rest of $\Sigma^{(2)}$ with $\ti{U}<0$ as the boundary.

\subsubsection*{Perturbation around vacuum}\label{smtexta}
\begin{figure}[ht]
  \centering
   \includegraphics[width=2in]{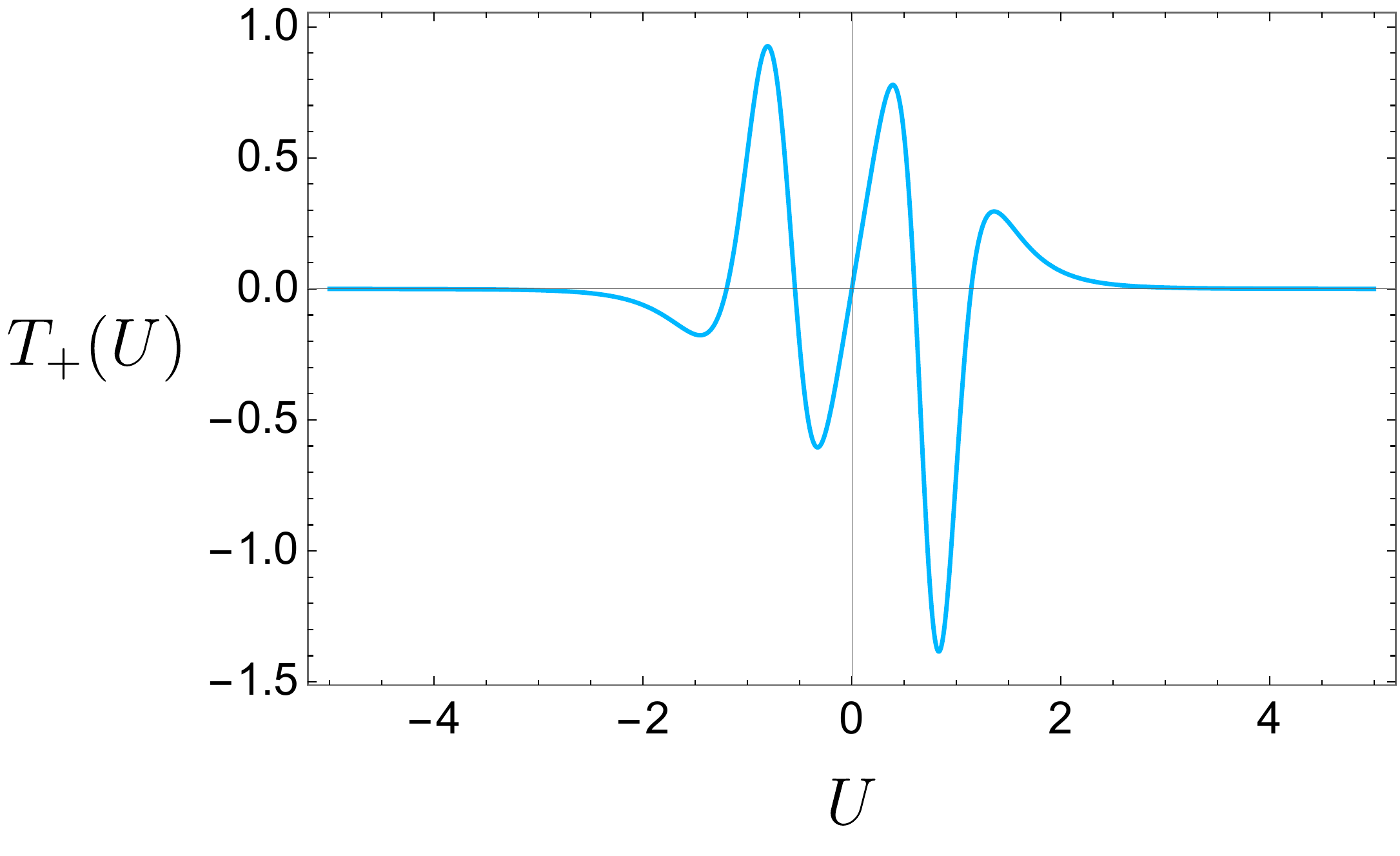}
    \includegraphics[width=2in]{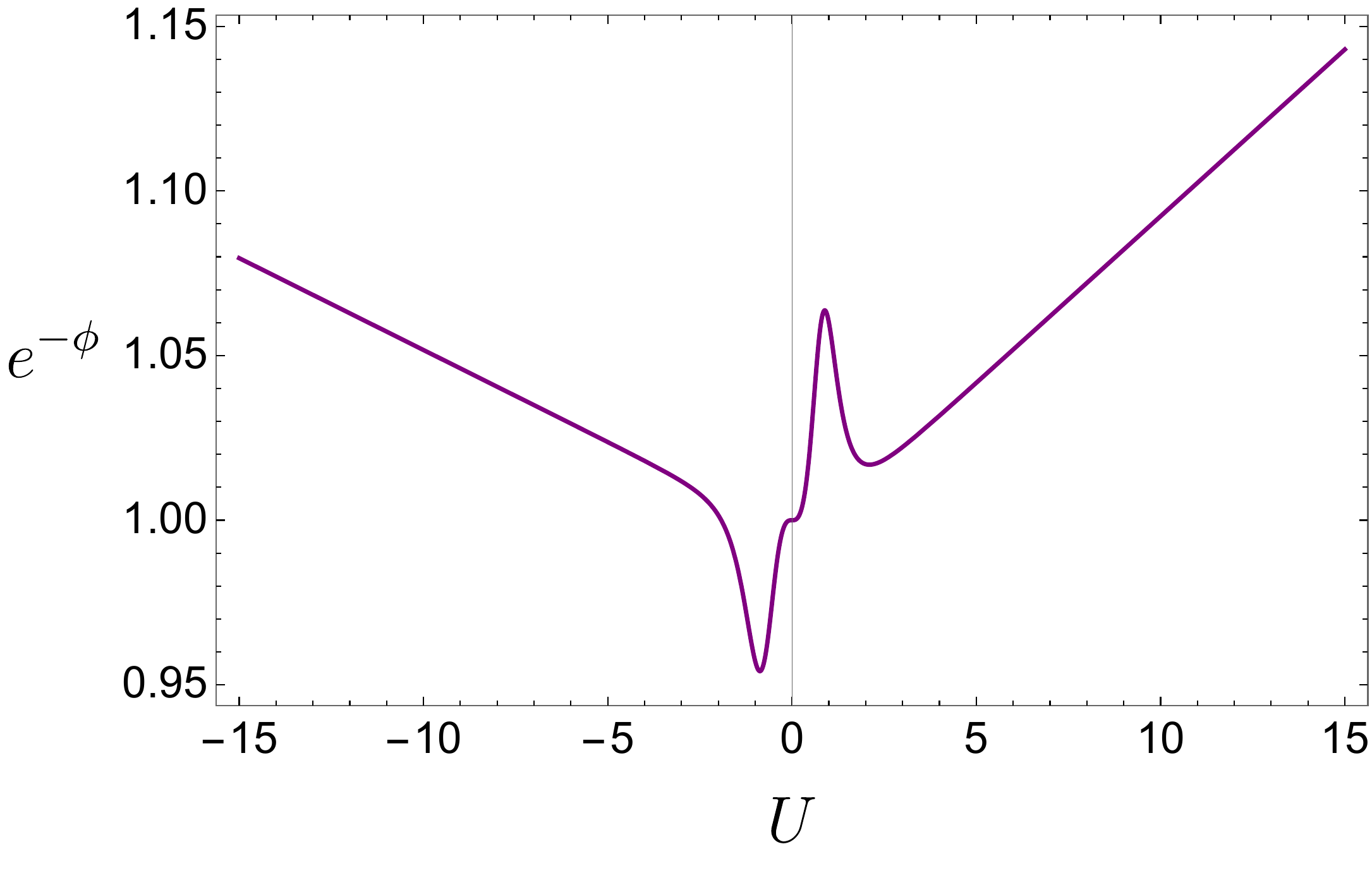}
     \includegraphics[width=2in]{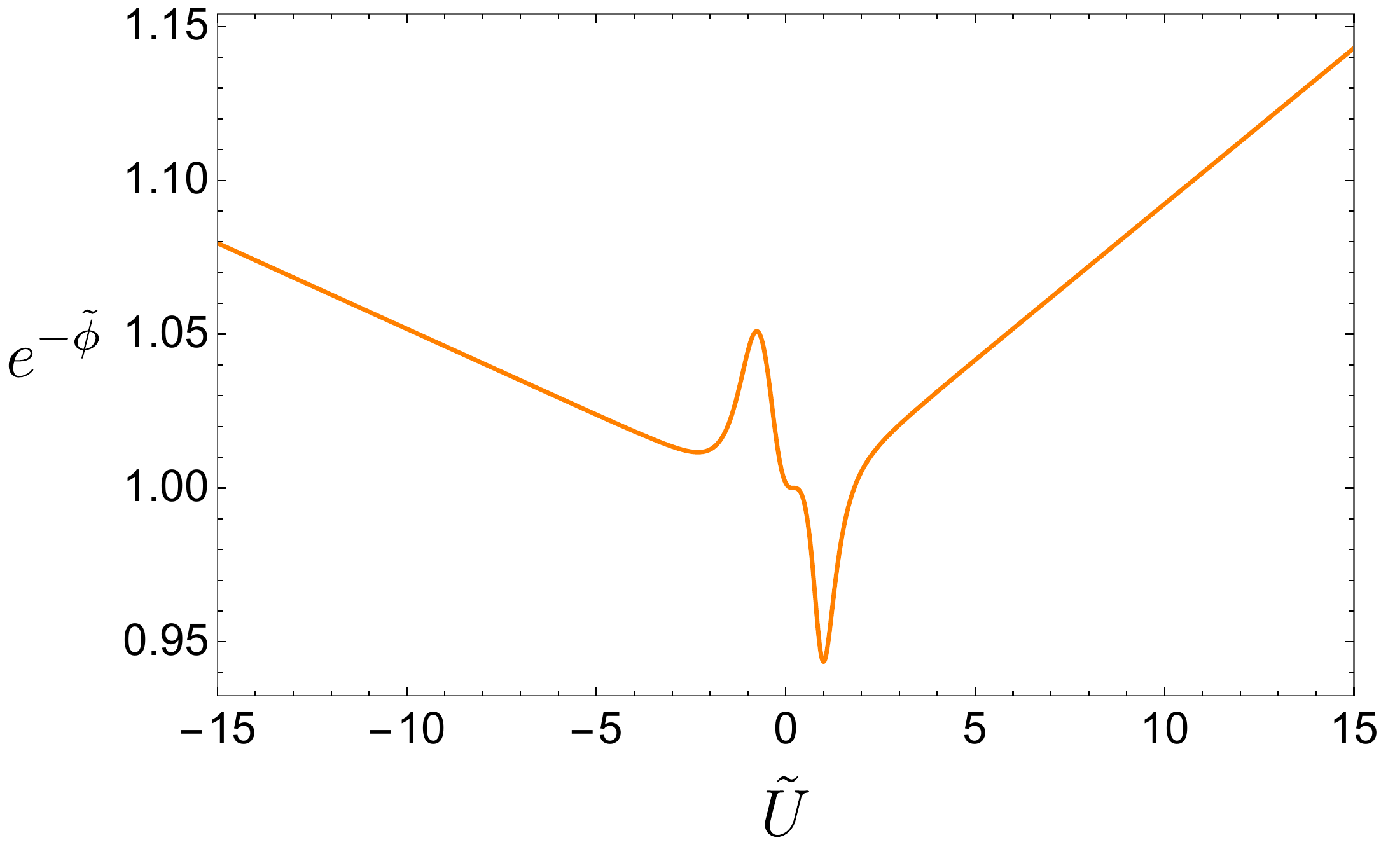}
  \caption{The left plot shows the function $T_+(U)$ as a function of null coordinate $U$. The middle and right plots describe the graphs of $e^{-\phi(U)}$
  and  $e^{-\ti{\phi}(\ti{U})}$, respectively. We have chosen $a=\frac{1}{5}$ for all plots.} 
\label{fig:Tuu}
\end{figure}

Finally, we introduce a function $P(U)$ that maps the real line $-\infty<U<\infty$ to $-\infty<\ti{U}<\infty$. For example, one explicit expression of $P(U)$ is given by 
\begin{equation}\label{puexp}
P(U)=U+\frac{a}{(1+U^4)}\,,
\end{equation}
with $a$ as a small constant. We will specifically use $a=1/5$ for our numerical calculations. The coordinate transformation $P(U)$ in eq.~\eqref{puexp} is a smooth and invertible function that plays the role of a source for generating smooth and non-trivial solutions. In the left panel of Figure~\ref{fig:Tuu}, we plot $T_+(U)$ as a function of $U$, which oscillates smoothly around zero. By solving the differential equations \eqref{eomphg} numerically, we can also compute the functions $e^{-\phi(U)}$ and $e^{-\ti{\phi}(\ti{U})}$, which are shown in the middle and right panels of Figure~\ref{fig:Tuu}, respectively. Asymptotically, it is straightforward to find that the solutions behave as
\begin{equation}
\phi(U)\simeq -\log |U|\,,
\end{equation}
as $U\to \pm\infty$, and similarly for $\ti{\phi}(\ti{U})$. We observe that the glued geometry, obtained by taking the union of the regions where $\eta\geq e^{-\phi(U)}$ and $\ti{\eta}\geq
e^{-\ti{\phi}(\ti{U})}$, is smooth and the two hypersurface $\Sigma^{(1)}$ and $\Sigma^{(2)}$ are glued together completely.

\subsection{Comments on energy condition}
\begin{figure}[t]
  \centering
   \includegraphics[width=2.3in]{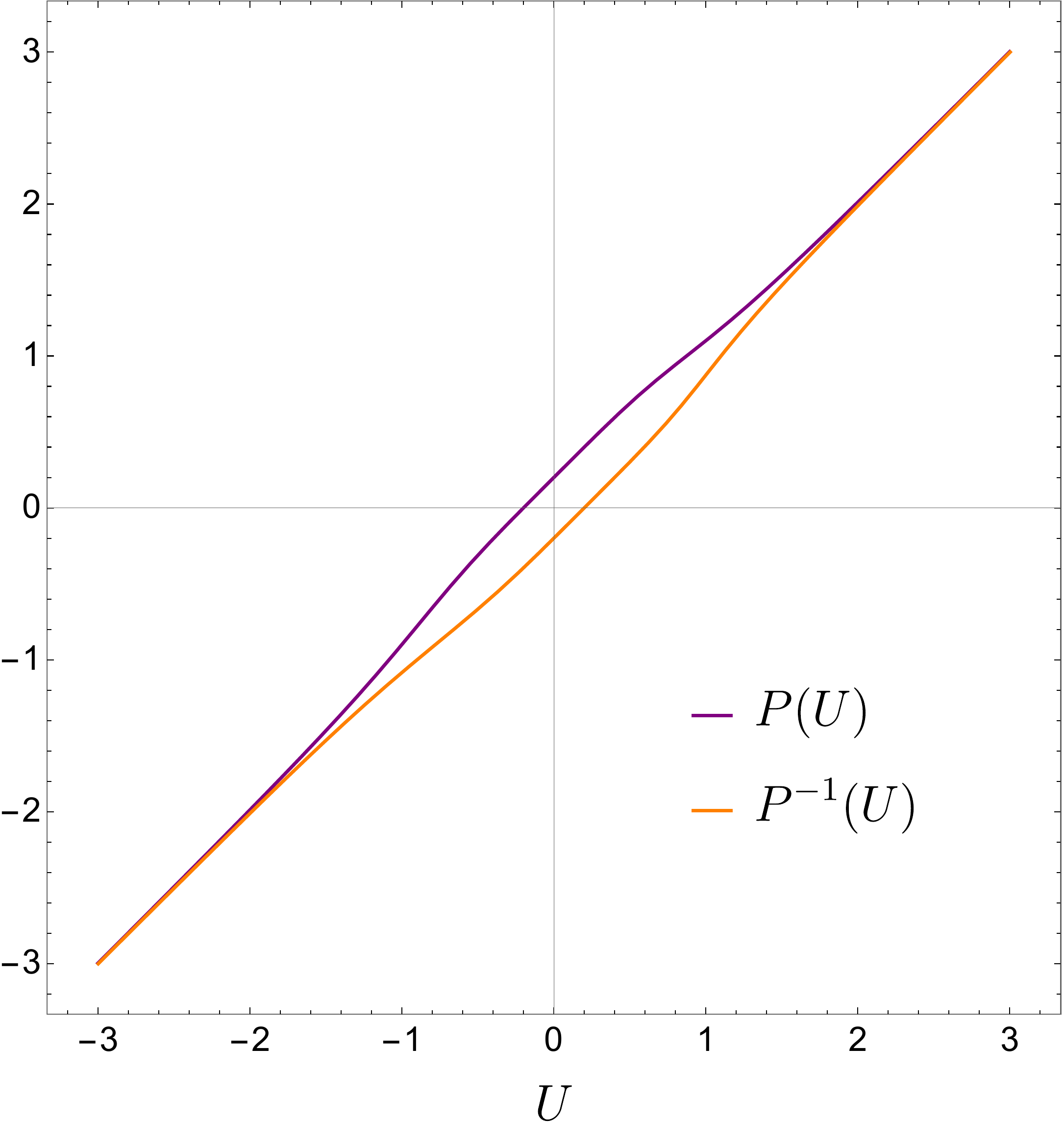}\qquad 
    \includegraphics[width=2.8in]{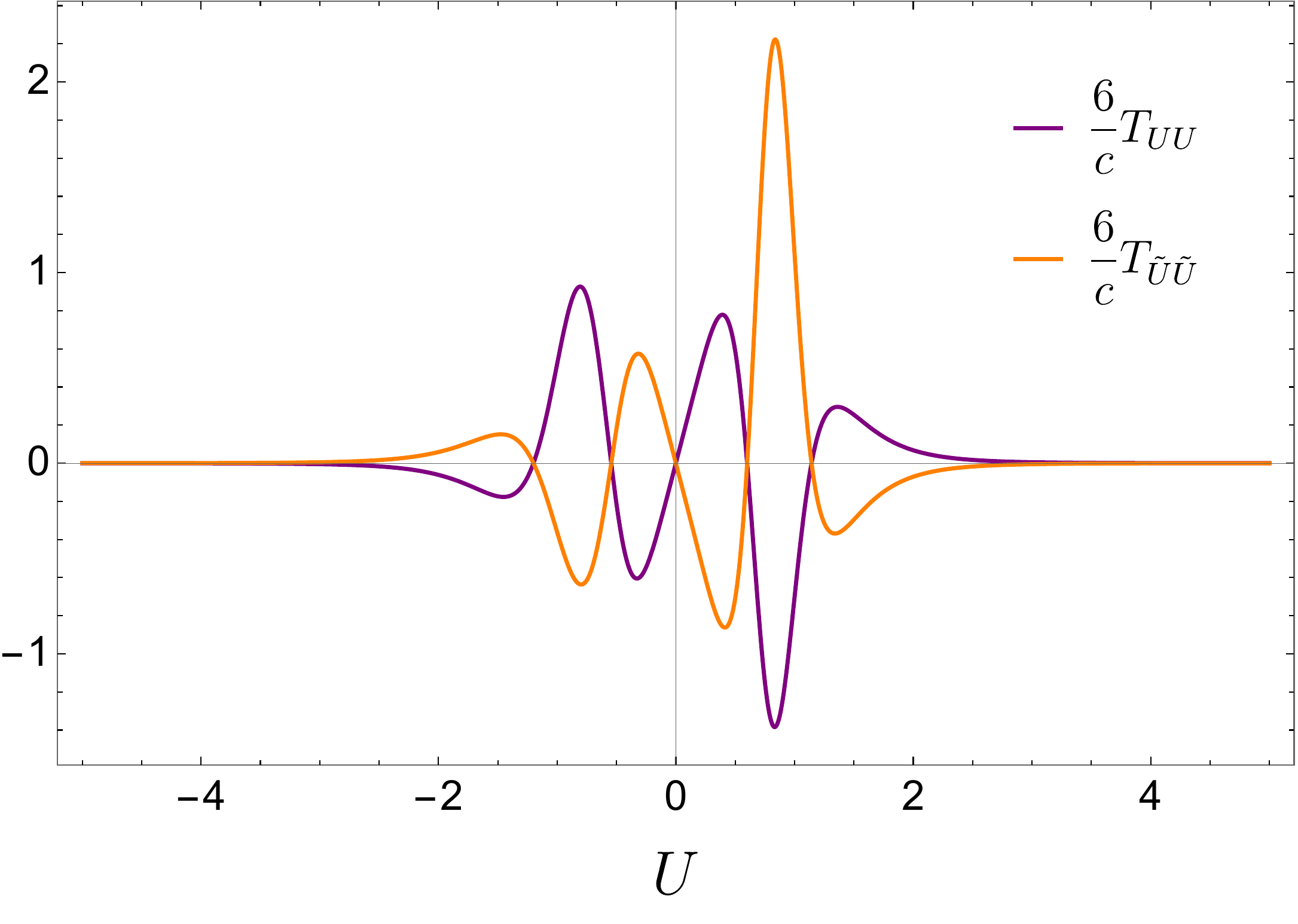}
  \caption{Left: function $P(U)$ given by eq.~\eqref{puexp} with taking $a=\frac{1}{5}$ and its inverse function $P^{-1}(U)$ as a function $U$. Right: energy stress tensors $T_{UU}(U)$ and $T_{\ti{U}\ti{U}}(P(U))$.
  } 
\label{fig:twoTuu}
\end{figure}

Because our gluing condition requires eq.~\eqref{eq:Einsteinbrane}, one might be concerned that one of the energy stress tensors among the two field theories will violate the energy condition. Here we would like to give a heuristic explanation of why this is not a problem. Consider an excited state in a two-dimensional CFT which is obtained by a conformal transformation $\ti{U}=P(U)$ from the vacuum state on a plane (described by $\ti{U}$). The energy stress tensor is computed by the conformal anomaly or the Schwarzian derivative:
\ba
T_{UU}=-\frac{c}{12}\{\ti{U},U\}=\frac{c}{6}
\left[\frac{3}{4}\left(\frac{P''(U)}{P'(U)}\right)^2-
\frac{P'''(U)}{2P'(U)}\right].\label{wmffgewww}
\ea
If we introduce $\Phi(U)$ such that
\ba
P'(U)=e^{-\Phi(U)},
\ea
then we find 
\ba
\frac{6}{c}T_{UU}=\frac{1}{2}\Phi''+\frac{1}{4}\Phi'^2.
\ea
This means that if we integrate the whole region $-\infty<u<\infty$, we find
\ba
\frac{6}{c}\int^\infty_{-\infty}dU T_{UU}=\left[\frac{1}{2}\Phi'\right]^{\infty}_{-\infty}+\int^\infty_{-\infty}dU \frac{\Phi'^2}{4}.
\ea
Thus, if we assume that $\Phi'$ gets vanishing in the limit $U\to \pm\infty$, which is the case when $p(U)$ approaches the vacuum value $p(U)=U$ in the limit
as in (\ref{puexp}), then we find
\ba
\frac{6}{c}\int^\infty_{-\infty}dU T_{UU}=\int^\infty_{-\infty}dU \frac{\Phi'^2}{4}\geq 0.
\ea
This is the averaged null energy condition (ANEC).
For the example of  (\ref{puexp}), we plotted this function and energy stress tensor in Figure \ref{fig:twoTuu}.

Then one may wonder if we can realize the condition like 
$T^{(1)}_{UU}+T^{(2)}_{\ti{U}\ti{U}}=0$, which is required by the gluing condition. Note that both $T^{(1)}_{UU}$ and $T^{(2)}_{\ti{U}\ti{U}}$ should satisfy the ANEC and do not seem to cancel each other. However, what we need to impose is the following condition:
\ba
T^{(1)}_{UU}+\left(\frac{d\ti{U}}{dU}\right)^2 T^{(2)}_{\ti{U}\ti{U}}=0\,. \label{gluecang}
\ea

Actually, this is satisfied by choosing the state of 
$T^{(2)}_{\ti{U}\ti{U}}$ such that it is obtained from the conformal transformation for the inverse map $U=P^{-1}(\ti{U})$, which leads to 
\ba
T^{(2)}_{\ti{U}\ti{U}}=-\frac{c}{12}\{U,\ti{U}\}=\frac{c}{12}\left(\frac{d\ti{U}}{dU}\right)^{-2}\cdot \{\ti{U},U\}.  \label{wpsjdwd}
\ea
Indeed, it is clear that this satisfies the condition (\ref{gluecang}).
It can also be seen that we have
\ba
&& \int^{\infty}_{-\infty}d\ti{U}
T_{\ti{U}\ti{U}}\geq 0\,, \no
&& \int^{\infty}_{-\infty}dU\left(\frac{d\ti{U}}{dU}\right)^{2}T_{\ti{U}\ti{U}}=-\int^{\infty}_{-\infty}dUT_{UU}\leq 0\,.
\ea

However, strictly speaking, we should note that in our gravity dual construction in section \ref{smtexta}, the coefficient of the Schwarzian derivative is halved as in eq.~\eqref{eq:tau12}.  Indeed, we cannot glue two solutions togehter with the stress tensors each given by eq.~\eqref{wmffgewww} and eq.~\eqref{wpsjdwd}. This is because the difference between the coordinates of $U$ in the first CFT and $\ti{U}$ in the second CFT looks like $\ti{U}=P(P(U))$ instead of $\ti{U}=P(U)$.  Thus, our gluing solution is not simply understood just as a standard conformal map. Nevertheless, the violation of ANEC is avoided in a similar way.

\subsection{Ba\~nados geometries}\label{sec:Banados}
One of the advantages of working on AdS$_3$ is that one can derive the most general vacuum solutions of Einstein equations with $\Lambda= - \frac{1}{L^2}$. By imposing Brown-Henneaux boundary conditions, one can find that the most general solutions of AdS$_3$ are given by \cite{Banados:1998gg}
\begin{equation}\label{eq:Banados}
    ds^2=\frac{dz^2}{z^2}+L^{(1)}_{+}(u)(du)^2+L^{(1)}_{-}(v)(dv)^2-\left(\frac{1}{z^2}
+z^2L^{(1)}_{+}(u) L^{(1)}_{-}(v)\right)dudv, 
\end{equation}
where $L^{(1)}_+(u), L^{(1)}_-(v)$ are arbitrary functions. They are the so-called Ba\~nados geometries. For example, the BTZ black hole corresponds to $L_+ +L_-= \frac{r_++r_-}{2}$ and $L_+- L_-=r_+r_-$. It is easy to check that the two arbitrary functions are nothing but the renormalized quasilocal stress tensors 
\begin{equation}
 \mathcal{T}_{ij} = -\frac{1}{8\pi \GN}\( K_{ij} - K h_{ij} + h_{ij}  \)  \Big|_{z\to 0} =  \frac{1}{8 \pi \GN}\begin{pmatrix}
 L^{(1)}_+ & 0 \\
 0 & L^{(1)}_-
\end{pmatrix}\,,
\end{equation}
 which are identified as holographic duals of boundary chiral and anti-chiral stress tensors.
Since AdS$_3$ geometries are locally the same, one can find coordinate transformations between two AdS$_3$ metrics. Beginning with the holographic dual of CFT$_2$ vacuum \ie AdS$_3$ in Poincar\'{e} metric 
\begin{equation}\label{eq:Poincare}
  ds^2=\frac{d\eta^2-dUdV}{\eta^2},   
\end{equation}
one can consider the conformal transformation on the boundary by taking 
\begin{equation}
U=p(u) \,, \qquad  V=q(v)\,.
\end{equation}
The corresponding bulk dual is given by coordinate transformations in AdS$_3$ as follows
\begin{equation}\label{eq:Banadosmap}
\begin{split}
U&=p(u)+\frac{2z^2(p')^2 q''}{4p'q'-z^2p''q''}\,,\\
V&=q(v)+\frac{2z^2(q')^2p''}{4p'q'-z^2p''q''}\,,\\
\eta&=\frac{4z(p'q')^{3/2}}{4p'q'-z^2p''q''}\,,
\end{split}
\end{equation}
which is known as Ba\~nados map 
\cite{Banados:1998gg,Roberts:2012aq,Shimaji:2018czt}. It is straightforward to check that the Poincar\'{e} metric eq.~\eqref{eq:Poincare} with this type of transformation is rewritten as the Ba\~nados metric defined in \eqref{eq:Banados} by identifying 
\begin{equation}
 L_+(u) = -\frac{1}{2} \{ p(u), u\} =\frac{3\left(p^{\prime \prime}\right)^2-2 p^{\prime} p^{\prime \prime \prime}}{4 p^2} \,, \quad   L_-(v) = -\frac{1}{2} \{ q(v), v\}= \frac{3\left(q^{\prime \prime}\right)^2-2 q^{\prime} q^{\prime \prime \prime}}{4 q^2}\,.
\end{equation}
In the following, we investigate the case by gluing two Ba\~nados geometries along a timelike brane. Similarly, we denote another bulk spacetime as 
\begin{equation}
   ds^2=\frac{d\tz^2}{\tz^2}+L^{(2)}_{+}(\ti{u})(d\ti{u})^2+L^{(2)}_{-}(\ti{v})(d\ti{v})^2-\left(\frac{1}{\ti{z}^2}
+\ti{z}^2L^{(2)}_{+}(\ti{u}) L^{(2)}_{-}(\ti{v})\right)d\ti{u}d\ti{v}\,, 
\end{equation}
which can be obtained from Poincar\'{e} metric \eqref{eq:Poincare02} by performing another conformal map $\tilde{U}=\ti{p}(\ti{u}), \tilde{V}=\ti{q}(\ti{v})$.

\subsubsection*{Finite Cut-off surface}
As we illustrated in the symmetric cases, the simplest solution of the junction conditions can be derived by taking $z=0=\tz$, $T=1$, and gluing arbitrary two Ba\~nados spacetimes along the conformal boundary. However, this is a very special case because the conformal boundary stays at the conformal infinity, where the energy flux is suppressed. To show how the constraint equation \eqref{eq:constrainAdS302} limits the possible configurations, we further consider gluing two Ba\~nados spacetimes on a finite cut-off surface located at  
\begin{equation}
	z = z_0\,, \qquad \tz= z_0 \,. 
\end{equation}
where we have chosen $z_0=\tz_0$ as the same constant due to the rescaling invariance of the bulk geometry (with rescaling the stress tensor $\tilde{L}_\pm$). Naively, the induced metric on the brane $\Sigma^{(1)}$ reads 
\begin{equation}
		ds^2 \big|_{\Sigma^{(1)}} = L_+(u)(du)^2+L_{-}(v)(dv)^2-\left(\frac{1}{z_0^2}
+z_0^2L_{+}(u) L_{-}(v)\right)dudv \,,\\ 
\end{equation}
An interesting observation is that this geometry is always flat regardless of the choices of $L_\pm$, which indicates that the first junction condition is naturally satisfied, \viz $\bar{R} [h^{(1)}] = 0 =\bar{R} [h^{(2)}]$. On the other hand, one can derive the corresponding extrinsic curvature by
\begin{equation}
 K^{(1)}_{uv}=	- \frac{1}{2 } \( \frac{1}{z_0^2}  - z_0^2 L_+ L_- \) \,,\qquad  K^{(1)}_{uu}=0=K^{(1)}_{vv}\,, 
\end{equation}
and 
\begin{equation}
   K^{(1)} = \frac{4}{ 1- z_0^4 L_+(u) L_-(v)} -2 \,.
\end{equation}
The traceless condition $K^{(1)}= 2T=K^{(2)} $ can be achieved by taking $T=1, z_0=0$ as one can expect. For a more general case with $z_0\ne 0$, the brane located at a finite cut-off $z=z_0$ only exists for \footnote{Of course, a special but different case is taking $L_+L_-=\text{constant}=\tilde{L}_+\tilde{L}_-$. However, this reduces to the symmetric cases discussed in section.~\ref{sec:symmetric}.}
\begin{equation}
	\begin{split}
		 T&= 1 \,, \\
		 L_+(u) L_-(v) &=0  =	 \tilde{L}_+(\tu) \tilde{L}_-(\tv) \,. 
		\end{split}
\end{equation}
In the following, let us choose $L_-(v)=0=\tilde{L}_-(\tv) $ without loss of generality. Indeed, this choice makes the disappearance of $T\bar{T}$ term explicit. For instance, one can evaluate the brane stress tensor and obtain 
\begin{equation}
\begin{split}
	\tau_{ij}^{(1)} &=\left(
	\begin{array}{cc}
		L_+ \left(T+\frac{4}{L_-  L_+ z_0^4-1}+2\right) & -\frac{(T+1) z_0^4 L_- L_+ +T+\frac{8}{z_0^4 L_-L_+ -1}+7}{2 z_0^2} \\
		\ast  & L_- \left(\frac{4}{L_- L_+ z_0^4-1}+T+2\right) \\
	\end{array}
	\right) 
	= \left( \begin{array}{cc}
		- L_+ (u) & 0 \\
		0   & 0\\
	\end{array}
	\right) \,,\\
 \tau_{ij}^{(2)} 
	&= \left( \begin{array}{cc}
		- \tilde{L}_+ (\tu) & 0 \\
		0   & 0\\
	\end{array}
	\right) \,, \\
 \end{split}
\end{equation} 
which is different from the trivial case with $\tau^{(a)}_{ij}=0$. It looks like we have found the possible solutions for gluing two branes $\Sigma^{(a)}$ at $z=z_0$ with any non-zero energy flux $L_+(u), \tilde{L}_+(\tu)$. Although we have shown the equivalence of the intrinsic geometry and the extrinsic geometry (\ie $K^{(1)}= 2T=K^{(2)}$) between the branes on two sides, we need to note that the existence of physical solutions (with real coordinates) implies more constraints. Recalling the original Israel junctions
\begin{equation}
\begin{split}
    L_+ (u) du^2 -\frac{1}{z_0^2} du dv &= 	 \tilde{L}_+ (\tu) d\tu^2 -\frac{1}{z_0^2} d\tu d\tv  \,, \\ 
    T_+(u){du}^2 + \tT_+ (\tu(u) )  d\tu^2&=0 \,, 
\end{split}
\end{equation}
it is obvious that the second junction condition can be solved if and only if 
\begin{equation}\label{eq:negative}
	 T_+(u) \tT_+(\tu) \le 0 \,.  
\end{equation}
Solving the Israel junction condition results in the connection between the two coordinate systems on the brane. Formally, one can recast the solutions as 
\begin{equation}
\begin{split}
    \tu (u)  &= \int  \sqrt{- \frac{L_+(u)}{\tilde{L}_+}}  d u  \,, \\
\end{split}	
\end{equation}
with assuming the satisfaction of eq.~\eqref{eq:negative}. More precisely, the transformation $\tu(u)$ can be solved by the following ODE
\begin{equation}
	L_+(u) + \tilde{L}_+ (\tu(u) )  \(  \frac{d \tu}{d u} \)^2 =0 \,. 
\end{equation}
The matching condition of the induced metric then leads us to another coordinate transformation, \viz
\begin{equation}
	\tv \( u, v \) =  \int    \sqrt{- \frac{\tilde{L}_+(\tu(u))}{L_+(u)}} \(  dv - 2 z_0^2 L_+(u) du   \)\,, 
\end{equation}
which is the formal solution for the following two PDEs:
\begin{equation}
	\frac{\partial \tv}{ \partial u} =  - 2 z_0^2 L_+  \sqrt{- \frac{\tilde{L}_+}{L_+}} \,, 
	\quad \frac{\partial \tv}{ \partial v} =    \sqrt{- \frac{\tilde{L}_+}{L_+}} \,. 
	\end{equation}

\subsection*{Chiral Solutions}
As we have shown, the possible solutions for gluing two Ba\~nados spacetimes are too restricted since we only consider the finite cut-off surface at $z=z_0$. To allow more general solutions as those chiral solutions discussed in the previous subsection for Poincar\'e AdS, we assume that the brane $\Sigma^{(1)}, \Sigma^{(2)}$ are located at
\begin{equation}\label{eq:chiralbrane02}
    z=e^{-\phi(u)} \,, \qquad \tilde{z}=e^{-\tilde{\phi}(\tilde{u})} \,,
\end{equation}
respectively. 

Different from the constant-$z$ slice, the intrinsic geometry of the brane at $z= e^{F(u)}$ gets more complicated. It is straightforward to obtain the induced metric at $\(u,v\)$ coordinates, namely
\begin{equation}
 		ds^2 \big|_{\Sigma^{(1)}} = \(L_+(u)+ \phi'^2\)(du)^2+L_{-}(v)(dv)^2-\left(e^{2\phi}
+e^{-2\phi}L_{+}(u) L_{-}(v)\right)dudv \,,\\ 
\end{equation}
whose Ricci scalar is expressed as 
\begin{equation}
	R[h^{(1)}]= \frac{8 e^{-6\phi} \phi'L_-' \(   2 (\phi')^2 -\phi''+e^{-4\phi} L_- ( 2L_+ (\phi')^2-L_+' \phi' +L_+\phi'')   \) }{ \(  1+ e^{-4\phi} \( e^{-4\phi} L_+^2 L_-^2  - 2 L_+ L_-  - 4L_- (\phi')^2  \)   \)^2 } \,. 
\end{equation}
It is clear that the flat brane $\Sigma^{(1)}$ (similar to $\Sigma^{(2)}$) only exists in two situations: 
\begin{equation}  
		\begin{cases}
		\phi'(u)=0\,, \text{with} \quad     z=\text{Constant} \,, \\
		L_-'(v)=0, \text{with} \quad  L_-=\text{Constant} \,.\\
	\end{cases}
\end{equation}
Since the first one has been explored before by taking the brane as a finite cut-off surface, we focus on the second case by setting $L_-(v)=L_-$ as a constant. On the other hand, the junction condition fixes the trace of the extrinsic curvature $K$, \ie 
\begin{equation}
\frac{2 (e^{-12\phi} L_+^3L_-^3 - e^{-8\phi} L_-^2 \(  L_+^2 +L_+ ( 8\phi'^2 +2\phi'' ) -2\phi'L_+'  \)  
 - e^{-4\phi}L_- ( L_+ +8\phi'^2 - 2\phi'')  +1  )}{ \(  1+ e^{-4\phi} \( e^{-4\phi} L_+^2 L_-^2  - 2 L_+ L_-  - 4L_- \phi'^2  \)  \)^{3/2}  } \,,
\end{equation}
to be a constant $2T$. With $T=1$, the simplest solution is given by $L_-=0$. Of course, As a generalization of the chiral solutions found in Poincar\'e AdS, the non-vanishing brane stress tensor associated with $\Sigma^{(1)}$in Ba\~nados geometry is given by  
\begin{equation}
 \tau_{uu}^{(1)}= \(\phi'(u)\)^2 - \phi''(u) - L_+(u)\,.
\end{equation}
with $\tau^{(1)ij}\tau^{(1)}_{ij}=0$. Similar to what we have shown before, one can explicitly find the coordinate transformation between $(u,v)$ and $(\tu, \tv)$ by solving the original Israel junction conditions. For example, the vanishing of $\tau^{(1)}_{uu}+\tau^{(2)}_{uu}$ fixes the relation between $u$ and $\tu$ as 
\begin{equation}
 L_+(u)-\(\phi'(u)\)^2+\phi''(u) +  \( \tilde{L}_+ (\tu(u) ) -\( \frac{d \tilde{\phi}}{d\tu}\)^2+\frac{d^2\tilde{\phi}}{d\tu^2}  \) \(  \frac{d \tu}{d u} \)^2 =0 \,.
\end{equation}
%%%%%%%%%%%%%%%%%%%%%%%%%%%%%%%%%%%%%%%%%%%%%%%%%%%%%%%%
%%%%%%%%%%%%%%%%%%%%%%%%%%%%%%%%%%%%%
\section{Non-chiral solutions for gluing AdS$_3/$CFT$_2$}
 \label{sec:Ngluing}
%%%%%%%%%%%%%%%%%%%%%%%%%%%%%%%%%%%%%%%%%%%%%%%%%%%%%%%%
%%%%%%%%%%%%%%%%%%%%%%%%%%%%%%%%%%%%%%%%%%%%%%%%%%%%%%%%
\subsection{Perturbative Construction}
In the preceding discussion, we focused on the special cases where the $T\bar{T}$ term vanishes. In these cases, the geometry of the brane is solely determined by the tension $T$, as seen in the brane constraint equation given by eq.~\eqref{eq:constrainAdS302}. However, in this section, we consider the effect of the $T\bar{T}$ term and study the curved brane geometry by gluing two Poincar\'e AdS$_3$ spacetimes whose line elements are defined by 
\begin{equation}\label{twoadss}
ds^2_{(1)}=\frac{d\eta^2-dUdV}{\eta^2}\,, \qquad ds^2_{(2)}=\frac{d\ti{\eta}^2-d\ti{U}d\ti{V}}{\ti{\eta}^2}\,.
\end{equation}
In the following, we still set $T=1$ as in previous sections. The most general ansatz for the brane positions $\Sigma^{(1)}$ and $\Sigma^{(2)}$ is given by
\begin{equation}\label{eq:gerenal surface in Poincare}
\eta=e^{-\phi(U,V)},\ \ \
\ti{\eta}= e^{-\ti{\phi}(\ti{U},\ti{V})}\,.
\end{equation}
The induced metric on $\Sigma^{(1)}$ is then obtained as
\begin{equation}\label{metawskl}
ds^2_{\Sigma}=-(e^{2\phi}-2\partial_U\phi\partial_V\phi)dUdV+(\partial_U\phi)^2dU^2
+(\partial_V\phi)^2dV^2\,,
\end{equation}
which is similar to that on $\Sigma^{(2)}$. The first Israel junction condition requires that the induced metrics on two sides of the brane, after gluing, should agree up to a coordinate transformation. Without loss of generality, we can assume the corresponding coordinate transformations are 
\begin{equation}
\ti{U}=A(U,V),\ \ \ \ \ti{V}=B(U,V)\,.
\end{equation}
On the other hand, the normal vector of $\Sigma^{(1)}$ as a hypersurface living in AdS$_3$ is obtained as 
\begin{equation}
(N^\eta,N^U,N^V)=-\frac{\eta}{\sqrt{1-4\eta^2\partial_U\phi\partial_V\phi}}
(1,-2\eta\partial_V\phi,-2\eta\partial_U\phi)\,,
\end{equation}
from which we can compute the extrinsic curvature.

While obtaining the most general brane profiles through the junction conditions is a formidable challenge, we can still make progress by exploring perturbative solutions for $\phi(U,V)$ and $\ti{\phi}(\ti{U},\ti{V})$. We can start from a finite cut-off surface located at $\eta=\eta_0$ and then construct solutions of $\phi(U,V)$ by taking the following series expansion: 
\begin{equation}
\phi(U,V)=\ep\cdot f(U,V)+\ep^2\cdot g(U,V)+\ep^3 \cdot h(U,V)+O(\ep^4)\,,
\end{equation}
with $\ep$ as a small parameter. Under this expansion, we can compute the scalar curvature $R^{(1)}$ and the trace of the extrinsic curvature $K^{(1)}$ on $\Sigma^{(1)}$ by 
\begin{equation}
\begin{split}
R^{(1)}&=8(\de_U\de_V f)\cdot \ep-8\left[2f(\de_U\de_V f)-(\de_U\de_V f)^2-\de_U \de_V g+(\de^2_U f)(\de^2_V f)\right]\ep^2+\ddd \,, \\
K^{(1)}&=2+4(\de_U\de_V f)\cdot \ep+\left(-8f\de_U\de_V f+4\de_U\de_V g)\right)\ep^2+ \cdots \,.
\end{split}
\end{equation}
Similar expressions $R^{(2)}$ and $K^{(2)}$ associated with $\Sigma^{(2)}$ can be also found.

Given the CMC condition $K^{(1)}=K^{(2)}=2$, as derived from eq.~\eqref{eq:K=2}, we need to set 
\begin{equation}
\de_U \de_V f=\de_U \de_V g=0 \,. 
\end{equation}
Using this information, we can express the functions $f(U,V)$ and $g(U,V)$ as
\begin{equation} \label{holfa}
\begin{split}
f(U,V)&=P(U)+Q(V)\,, \quad g(U,V)=\ap(U)+\beta(V),\\
\ti{f}(\ti{U},\ti{V})&=\ti{P}(\ti{U})+\ti{Q}(\ti{V})\,,\quad \ti{g}(\ti{U},\ti{V})=\ti{\ap}(\ti{U})+\ti{\beta}(\ti{V})\,,
\end{split}
\end{equation}
where $P(U), Q(V), \ap(U)$, and $\beta(V)$ are arbitrary functions.
By imposing $K^{(1)}=2$ again, we obtain the relation $\de_U \de_V h =-(Q'^2 P'' + P'^2 Q'')$. Assuming the form of solutions given in (\ref{holfa}), we can simplify the expressions of $R$ and $K$ up to the order of $O(\ep^3)$ as follows:
\begin{equation*}
\begin{split}
    R^{(1)} &=-8P''Q''\cdot \ep^2+8\left[(Q')^2P''+(P')^2 Q''+4(P+Q)P''Q''-Q''\ap''-P''\beta''\right]\ep^3+O(\ep^4)\,,\\
K^{(1)}&=2+O(\ep^4)\,,
\end{split}
\end{equation*}
Moreover, the brane stress tensor $\tau^{(1)}_{ij}=K^{(1)}_{ij}-K^{(1)}h^{(1)}_{ij}+h^{(1)}_{ij}$ can be derived as
\begin{equation}\label{pertstgh}
  \begin{split}
       \tau^{(1)}_{UU} &=-P''\cdot \ep+\left[(P')^2-\ap''\right]\ep^2
+\left[2P'\ap'-2P'Q'P''-\de^2_U h\right]\ep^3+O(\ep^4)\,,\\
 \tau^{(1)}_{VV} &=-Q''\cdot \ep+\left[(Q')^2-\beta''\right]\ep^2
+\left[2Q'\beta'-2Q'P'Q''-\de^2_V h\right]\ep^3+O(\ep^4)\,,\\
 \tau^{(1)}_{UV}&=\left[(Q')^2P''+(P')^2Q''\right]\ep^3+O(\ep^4)\,.
  \end{split}
\end{equation}
The validity of the Israel junction condition necessitates the following relations:
\begin{equation}\label{EMcondhew}
\tau^{(1)}_{UU}+\left(\frac{d\ti{U}}{dU}\right)^2 \tau^{(2)}_{\ti{U}\ti{U}} =0\,,\quad \text{and} \quad
\tau^{(1)}_{VV}+\left(\frac{d\ti{V}}{dV}\right)^2 \tau^{(2)})_{\ti{V}\ti{V}}=0\,.
\end{equation}
Similar expressions and relations can be derived for the brane stress tensor $\tau^{(2)}_{\ti{U}\ti{U}}$ and $\tau^{(2)}_{\ti{V}\ti{V}}$ in the second BFT.
It is important to note that up to $O(\ep^2)$, we have $\tau^{(1,2)}_{UV}=0$, as is evident from eq.~\eqref{pertstgh}. Consequently, the junction condition is automatically satisfied for the $(UV)$ component.

To obtain the explicit solutions, we begin by considering the relation between the $(U,V)$ and $(\ti{U},\ti{V})$ coordinates. With equating the first induced metric \eqref{metawskl} and the second one at the leading order, we can get
\begin{equation}
    \begin{split}
     \ti{U}=A(U)=U+A_1(U)\ep+A_2(U)+O(\ep^3)\,,\\ \ti{V}=B(V)=V+B_1(V)\ep+B_2(V)+O(\ep^3)\,,
\end{split}
\end{equation}
with $A_1'(U)=4P(U)$ and $B_1'(V)=4Q(V)$. Next, we solve the junction condition \eqref{EMcondhew}. At the order of $O(\ep)$, we have
\begin{equation}
\ti{P}(\ti{U})|_{\ti{U}=A(U)}=-P(U)\,, \qquad 
\ti{Q}(\ti{V})|_{\ti{V}=B(V)}=-Q(V)\,.
\end{equation}
At the next order $O(\ep^2)$, the condition is solved by 
\begin{equation}
 \begin{split}
    \ap''+\ti{\ap}(U)''&=-2(P')^2\,, \qquad \beta''+\ti{\beta}(V)''=-2(Q')^2\,,\\
    -\alpha+\ti{\alpha}-4P^2+\frac{1}{2}A_2'&=0\,, \qquad -\beta +\ti{\beta}-4Q^2+\frac{1}{2}B_2''(V)=0.
\end{split}   
\end{equation}
Using these solutions, one can explicitly show 
\begin{equation}
\begin{split}
\tau^{(1)ij}h_{ij} &=O(\ep^4)\,,\\
 R+\tau^{(1)ij}\tau^{(1)}_{ij}&=O(\ep^4)\,,\\
\end{split}
\end{equation}
where $R^{(1)}$ and $R^{(2)}$ are identical and thus denoted simply as $R$. This matches with eq.~\eqref{eq:constrainAdS3} obtained from the general analysis.
Therefore, the above solutions provide a class of perturbative solutions with non-chiral excitations.
%%%%%%%%%%%%%%%%%%%%%%%%%%%%%%%%%%%%%%%%%%%%%%%%%%%%%%%%
%%%%%%%%%%%%%%%%%%%%%%%%%%%%%%%%%%%%%
\section{Another approach based on wedge holography}
 \label{sec:Wedge}
%%%%%%%%%%%%%%%%%%%%%%%%%%%%%%%%%%%%%%%%%%%%%%%%%%%%%%%%
%%%%%%%%%%%%%%%%%%%%%%%%%%%%%%%%%%%%%%%%%%%%%%%%%%%%%%%
\begin{figure}[t!]
  \centering
   \includegraphics[width=8cm]{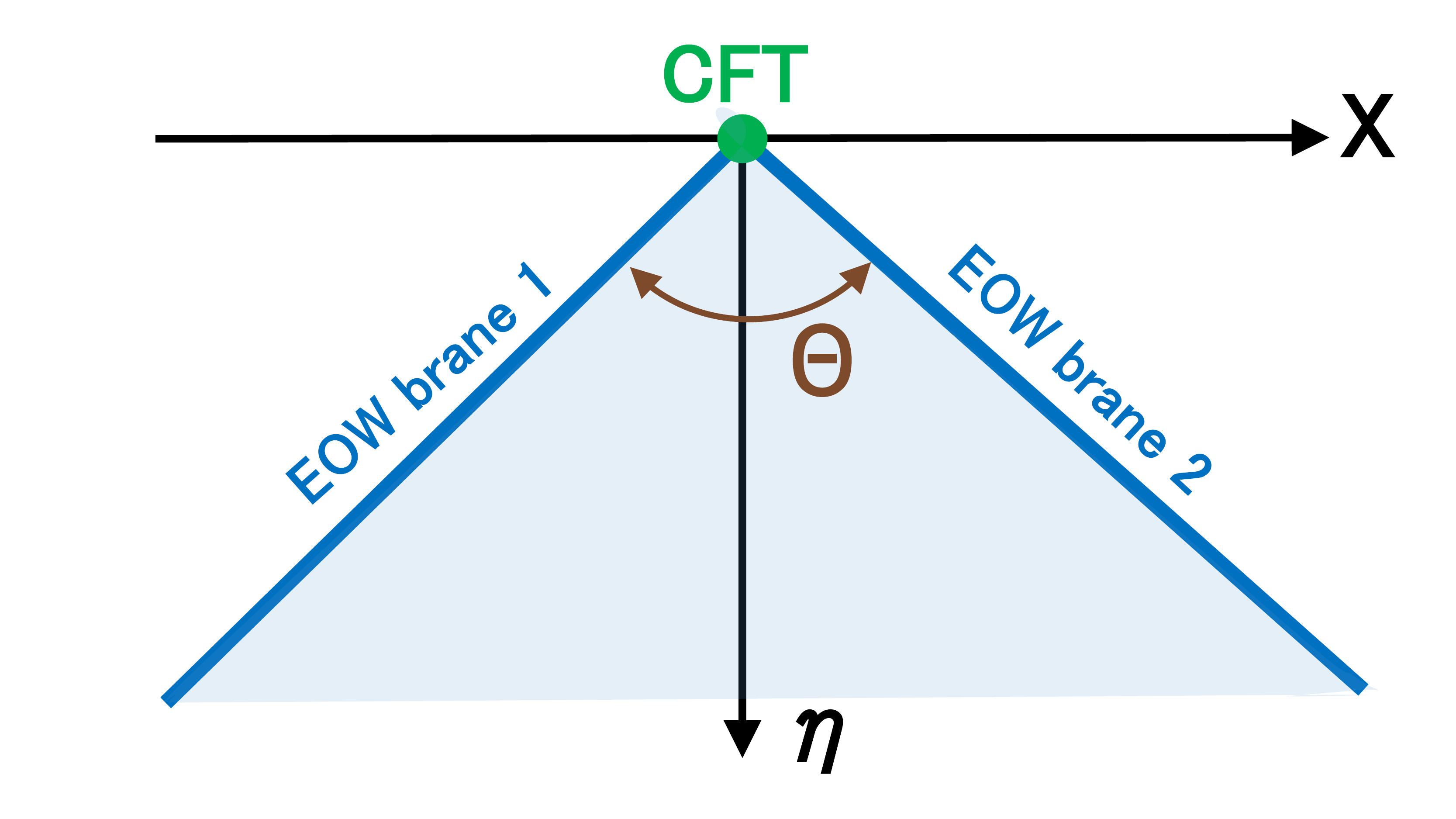}
  \caption{Sketch of Wedge Holography. The wedge region (blue colored region) is surrounded by two EOW branes (blue lines). The intersection of two EOW branes is the tip of the wedge (green dot), where the dual CFT lives.}
\label{fig:glueAdS}
\end{figure}

Before we conclude this paper, we would like to briefly discuss another method for gluing AdS/CFT. This is to employ wedge holography \cite{Akal:2020wfl}. As depicted in Figure \ref{fig:glueAdS}, we consider a $d+1$-dimensional wedge-like region in Poincar\'e metric 
\begin{equation}
ds^2=\frac{-dt^2+dx^2+d\eta^2+\sum_{i=1}^{d-2}dx^2_i}{\eta^2}\,.
\end{equation}
The wedge region is surrounded by two EOW branes, where we impose the Neumann boundary condition with a constant value of tension.
The wedge holography states the chain of duality, which first argues that the gravity on the $d+1$-dimensional wedge region is dual to the $d$-dimensional quantum gravity on the EOW branes. Secondly, this gravity is dual to a $d-1$ dimensional CFT on the tip of the wedge. The intermediate picture in $d$-dimension looks identical to our setup of gluing two AdS geometries. In the original wedge holography, we impose the Dirichlet boundary condition on the tip. However, for our proposal of gluing AdS/CFT, we need to impose the Neumann boundary condition on the tip, which is equivalent to fixing the angle $\theta$ of the intersection of two EOW branes.

Below, we will focus on the case where the bulk spacetime is part of AdS$_3$ with three-dimensional pure gravity. When the bulk metric is given by the Poincar\'e metric \eqref{pointh}, the simplest profile of EOW branes takes the form $X=\lambda_1 \eta$ and $X=\lambda_2\eta$ as depicted in Figure \ref{fig:glueAdS}. This corresponds to the vacuum solution of gluing AdS/CFT. 

\begin{figure}[t!]
  \centering
  \includegraphics[width=7cm]{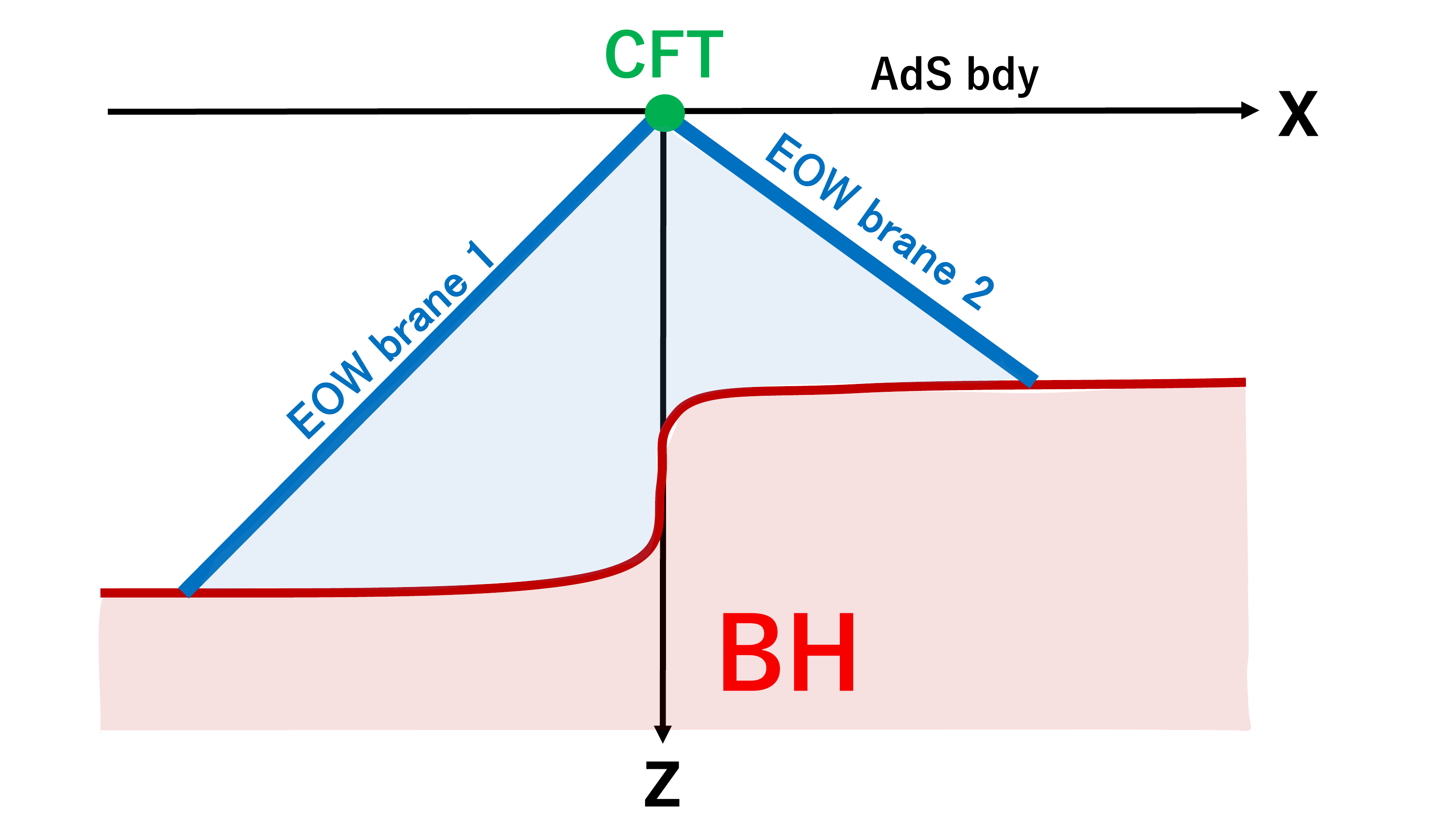}
  \includegraphics[width=7cm]{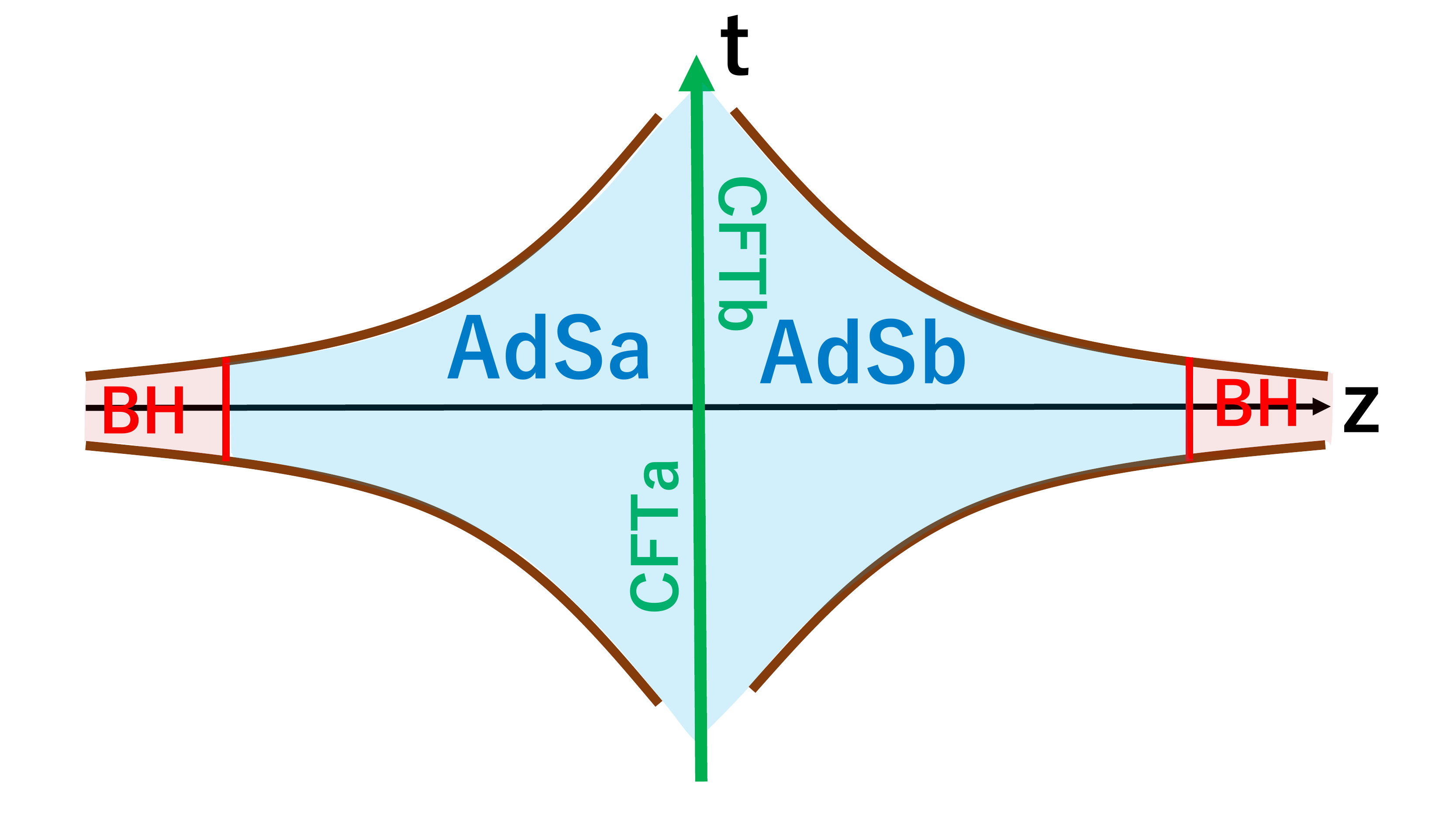}
  \caption{Wedge holography in the presence of the black hole horizon (left) and its boundary dual which describes gluing two AdS black hole solutions (right).}
\label{fig:glueEOW}
\end{figure}

To describe non-vacuum solutions, we can introduce a black hole in the bulk, as in the left panel of Figure \ref{fig:glueEOW}. This is dual to gluing two AdS black hole geometries together, as depicted in the right panel of Figure \ref{fig:glueEOW}. We can even create a situation in which two AdS$_2$ geometries with different temperatures. In such a bulk solution, the temperature of the black hole at $x=-\infty$ is different from that at $x=\infty$. This can be found by considering the gravity dual of the following conformal map.
\ba
\zeta = e^{\frac{2\pi}{\beta}z},
\ea
and
\ba
w=z+\frac{\ap}{2\pi}\log\left[\cosh\left(\frac{2\pi}{\beta}z\right)\right].
\ea
The first transformation maps a half plane Re$\zeta>0$ into a cylinder, which leads to a state at the inverse temperature $\beta$. The second one maps the cylinder into an inhomogeneous one. Note that this treatment is a special example of inhomogeneous quantum quenches  \cite{Sotiriadis:2008ila}.

The coordinates in the Lorentzian signature can be obtained from the following Wick rotation:
\ba
&& (U,V)=(T-X,T+X)=(-\zeta,\bar{\zeta})\,,\no
&& (u,v)=(t-x,t+x)=(-w,\bar{w})\,.
\ea
The coordinate transformations are thus derived as 
\ba
&& -u=\frac{\beta}{2\pi}\log\left(-U\right)
+\frac{\ap}{2\pi}\log\left[\frac{-U-1/U}{2}\right],\no
&& v=\frac{\beta}{2\pi}\log V
+\frac{\ap}{2\pi}\log\left[\frac{V+1/V}{2}\right].\label{ewhso}
\ea
This shows that the state, described by the coordinates $(u,v)$ has the inverse temperature $\beta\pm\ap$ in the limit $x\to \pm\infty$. We can find the metric of the inhomogeneous black hole solutions by plugging the above transformations into eq.~\eqref{eq:Banadosmap}
and deriving the Ba\~nados metric \eqref{eq:Banados}. The EOW branes located at $x=\lambda_{1,2} \eta$ in Poincar\'e AdS$_3$ are also mapped into those in the Ba\~nados geometry via eq.~\eqref{eq:Banadosmap}. Thus we obtain the bulk solution of the wedge holography depicted in the left panel of Figure \ref{fig:glueEOW}. 

In this way, wedge holography provides another useful method for finding solutions for gluing AdS/CFT, at least for two-dimensional gravity, albeit through an indirect method utilizing the holography. One may wonder why we can find the above solution by gluing two AdS black holes, which was missing in our direct analysis of gluing AdS/CFT in the previous sections. However, we need to note that we imposed the Neumann boundary condition at the tip of the wedge, which is expected to correspond to the junction condition \eqref{eq:Israel}. Since the tip is situated at the strict AdS boundary $\eta=0$, the gravity back-reaction at the glued surface (called $\Sigma$ in previous sections) is negligible. On the other hand, in the previous sections, we took into account the dynamical gravity and considered the generic situations where $\Sigma$ is located at finite $\eta$. Indeed, even in our wedge holographic construction, if we choose the intersection of two EOW branes to be located at finite $\eta$, the intersection would get more complicated, where the intersecting angle $\theta$ would become position dependent in general. This no longer satisfies the Neumann boundary condition, which requires a constant value of $\theta$. Instead, this can be a solution only if we appropriately arrange the matter energy stress tensor at the intersection $\Sigma$ so that it solves the junction conditions.

%%%%%%%%%%%%%%%%%%%%%%%%%%%%%%%%%%%%%%%%%%%%%%%%%%%%%%%%
%%%%%%%%%%%%%%%%%%%%%%%%%%%%%%%%%%%%%
\section{Discussions}\label{sec:Conclusion}
%%%%%%%%%%%%%%%%%%%%%%%%%%%%%%%%%%%%%%%%%%%%%%%%%%%%%%%%
%%%%%%%%%%%%%%%%%%%%%%%%%%%%%%%%%%%%%%%%%%%%%%%%%%%%%%%
In this paper, we consider gluing two AdS spacetimes by using a timelike brane with constant tension to construct a non-boundary holographic spacetime, which is different from the standard AdS/CFT. The gluing between the two sides is realized by performing the Israel junction conditions \eqref{eq:Israel}. We first show in eq.~\eqref{eq:Kconstant} that the junction conditions guarantee that the brane with respect to each side is always given by a constant mean curvature slice whose trace of the extrinsic curvature is determined by the tension of the brane. Despite these geometric constraints, we would like to interpret the junction condition as the ``Einstein equation" on the brane, \ie eq.~\eqref{eq:Einsteinbrane} with respect to its induced metric. As a result of the CMC condition, the brane stress tensors are always fixed to be traceless. Using the Gauss equation for the codimension-one brane, we show in eq.~\eqref{eq:constrainAdS302} that the intrinsic curvature of the brane geometry is controlled by the $T\bar{T}$ term of the brane stress tensor, which differs from standard Einstein gravity. We focus on the special cases in the rest of the paper by gluing two AdS$_3$ along a two-dimensional brane. In particular, we present solutions of various types of brane profile by considering Poincar\'e AdS$_3$, Ba\~nados geometries, and including nonvanishing brane stress tensors. 

\subsection*{Effective brane theory} 
Given that the brane truncates the bulk spacetime on either side, it is plausible to consider the joint AdS spacetime as a non-boundary bulk spacetime. Nonetheless, it is reasonable to expect that a holographic effective theory exists on the brane that captures the dynamical degrees of freedom of the bulk spacetime. Prior to the gluing of the two bulk spacetimes, specifically along a generic timelike brane $\Sigma^{(a)}$, which is regarded as a finite cut-off surface, it is known that the corresponding boundary theory is defined by a $T\bar{T}$ deformed CFT. Within the context of two-dimensional brane field theory, the act of gluing the two bulk spacetimes along the brane corresponds to the coupling of the two field theories residing on $\Sigma^{(1)}$ and $\Sigma^{(2)}$, given that the Dirichlet boundary condition is deactivated and the two brane field theories interact by virtue of the induced gravity on the brane. A pivotal question is: what is this interacting brane field theory? In principle, the effective action of the brane field theory can be derived from the gravitational action, \ie 
\begin{equation}
    I_{\mt{BFT}} \equiv I_{\mathrm{total}}=I_{\mathrm{bulk}}+I_{\mathrm{bdy} }\,, 
\end{equation}
by integrating each side to the position of the brane.

The nature of the interacting brane field theory remains uncertain for a generic brane profile. However, when the brane is taken to the conformal boundary, it can be shown that the brane field theory is a sum of two Liouville field theories.
This can be established by parametrizing the regular intrinsic metric of the brane as an off-diagonal form $\gamma_{ij}dx^i dx^j = - e^{2\Phi(u,v)}du dv$ and performing the limit $\epsilon \to 0$ that takes the brane at $z= \epsilon e^{- \Phi}$ to the  conformal boundary. In this limit, the effective action reduces to the Liouville field theory, as has been demonstrated in the literature, see \eg \cite{Carlip:2005tz,Carlip:2005zn,Nguyen:2021pdz} for more details. 
 Specifically, the effective action is derived as 
\begin{equation}
 \lim_{\epsilon \to 0} I_{\mt{BFT}} \approx \frac{1}{16 \pi \GN} \int d^2 x \sqrt{|\gamma|}\left(\nabla^i \Phi\nabla_i \Phi + \Phi R [\gamma]\right)  + \tilde{\Phi}\,\,\text{part}\,,
\end{equation}
with an additional part given by $\tilde{\Phi}$ from $\Sigma^{(2)}$. 
The energy-stress tensor of the Liouville field $\Phi$ is obtained from the effective action as
\begin{equation}
T_{ij}^{(1)}  \propto  \left( \begin{array}{cc}
	\(\partial_u \Phi \)^2-  \partial_u^2 \Phi  & - \partial_u \partial_v \Phi   \\
	 - \partial_u \partial_v \Phi  &  \(\partial_v \Phi \)^2-  \partial_v^2 \Phi  \\ 
\end{array}
\right) \,.
\end{equation}
The Einstein equation on the brane leads to the Israel junction condition, which states that the sum of the energy-stress tensor of the two Liouville fields must vanish, \ie 
\begin{equation}
    \frac{\delta I_{\mt{BFT}} }{\delta \gamma^{ij}} =0  \quad \rightarrow \quad   T_{ij}^{(1)} +T_{ij}^{(2)}=0 \,. 
\end{equation}
On the other hand, the equation of motion for the Liouville field $\Phi$: 
\begin{equation}
    \frac{\delta I_{\mt{BFT}} }{\delta \Phi} \rightarrow   \quad   \nabla^{2} \Phi =- \frac{1}{2} R[\gamma] \,,
\end{equation}
indicates $\partial_u \partial_v \Phi=0 $ for on-shell solutions on a flat conformal boundary such as that in Ba\~nados geometry. For example, we can parametrize the on-shell solutions as $\Phi (u,v)=\phi(u) + \frac{1}{2} \log( \frac{p'(u)}{(1-p(u))^2})$, which exactly produces our previous result, \ie $T^{(1)}_{uu} \propto  (\phi')^2 - \phi'' - L_+ (u)$ as shown in section \ref{sec:Banados} for gluing two Ba\~nados geometries.

Certainly, the examination presented herein is restricted to the particular scenario in which the brane is located at the conformal boundary. However, in the context of a more generic brane living in bulk spacetime, it is reasonable to anticipate that the Liouville field theories would be deformed by a $T\bar{T}$ term, and interact with each other. From the perspective of two-dimensional holographic BFTs, it is reasonable to anticipate that the total Hamiltonian can be expressed as $H=H^{(1)}+H^{(2)}+H_{\mathrm{int}}$. The specific form of the interaction term $H_{\mathrm{int}}$ can be understood in terms of the $T^{(1)}T^{(2)}$ deformation, which arises due to the exchange of gravitons between two AdS spacetimes. This kind of deformation, associated with the $T^{(1)}T^{(2)}$ term, has been examined in the context of conventional CFTs in \cite{Ferko:2022dpg}. Also, it is intriguing to note that the condition \eqref{eq:Einsteinbrane} implies the total central charge is vanishing. This is what we expect when we couple a CFT with two-dimensional gravity. Even if the original CFT has a positive central charge, the Liouville CFT, which emerges from the diagonal metric fluctuations of gravity, has a negative central charge that cancels the original one and results in a vanishing total central charge.

\subsection*{Open quantum system}
Instead of treating the two portions of AdS spacetime on equal footing, an alternative approach is to regard one of the bulk spacetimes as the environment with respect to another one. This strategy is commonly employed in the theory of open quantum systems, where a target system and its surrounding environment are considered to be two distinct systems. By taking the partial trace over the degrees of freedom of the environment, one can obtain the non-unitary time evolution of the target system \cite{open}. In the context of our gluing AdS/CFT setup, one can identify one of the AdS spacetimes as the environment and the brane as the interface between the target system and the environment. The joint spacetime then realizes a holographic realization of the open quantum system.

\subsection*{Gluing two de Sitter spacetimes} 
\begin{figure}[ht]
  \centering
   \includegraphics[width=2in]{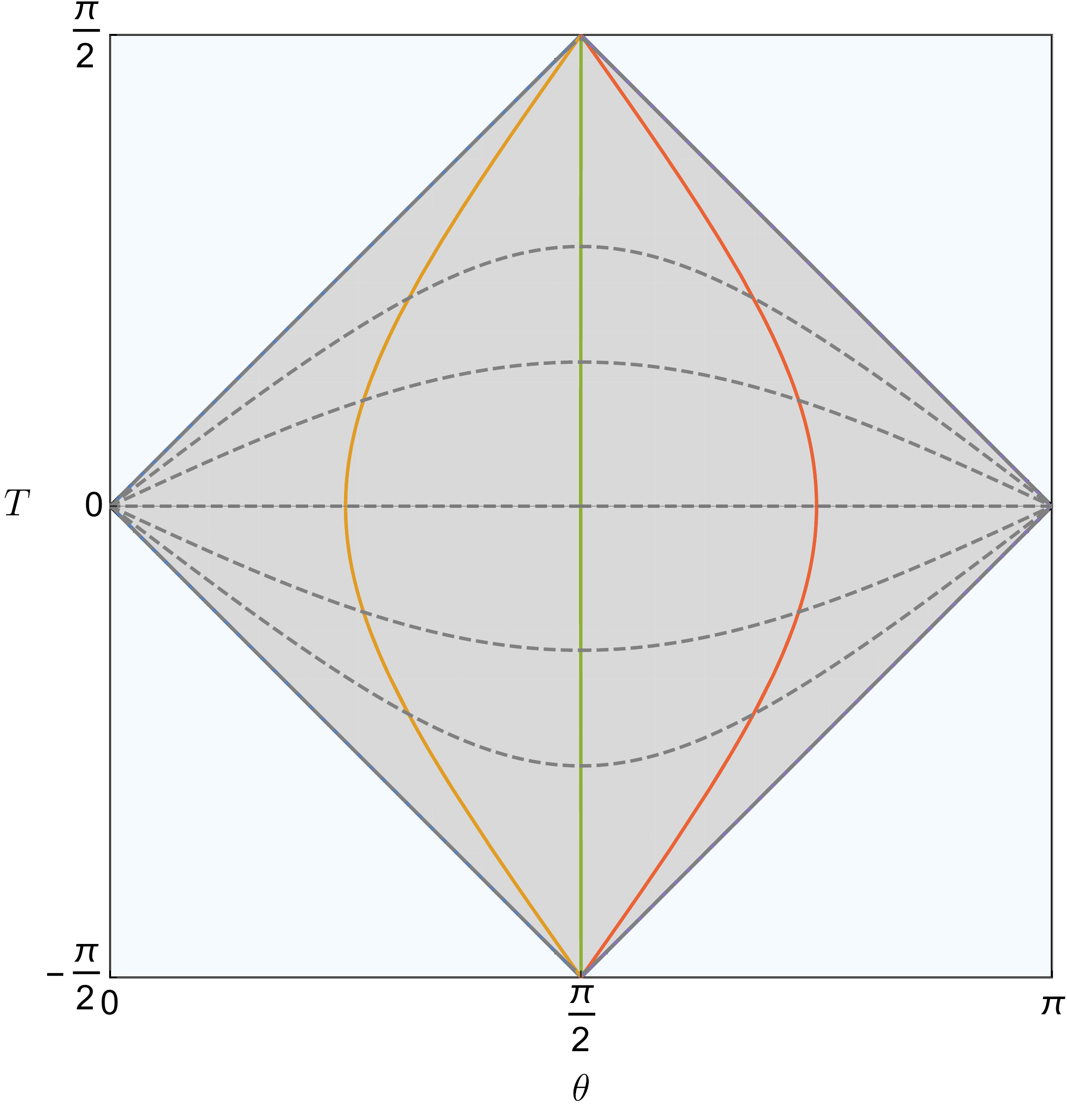}
   \quad \quad 
    \includegraphics[width=2.3in]{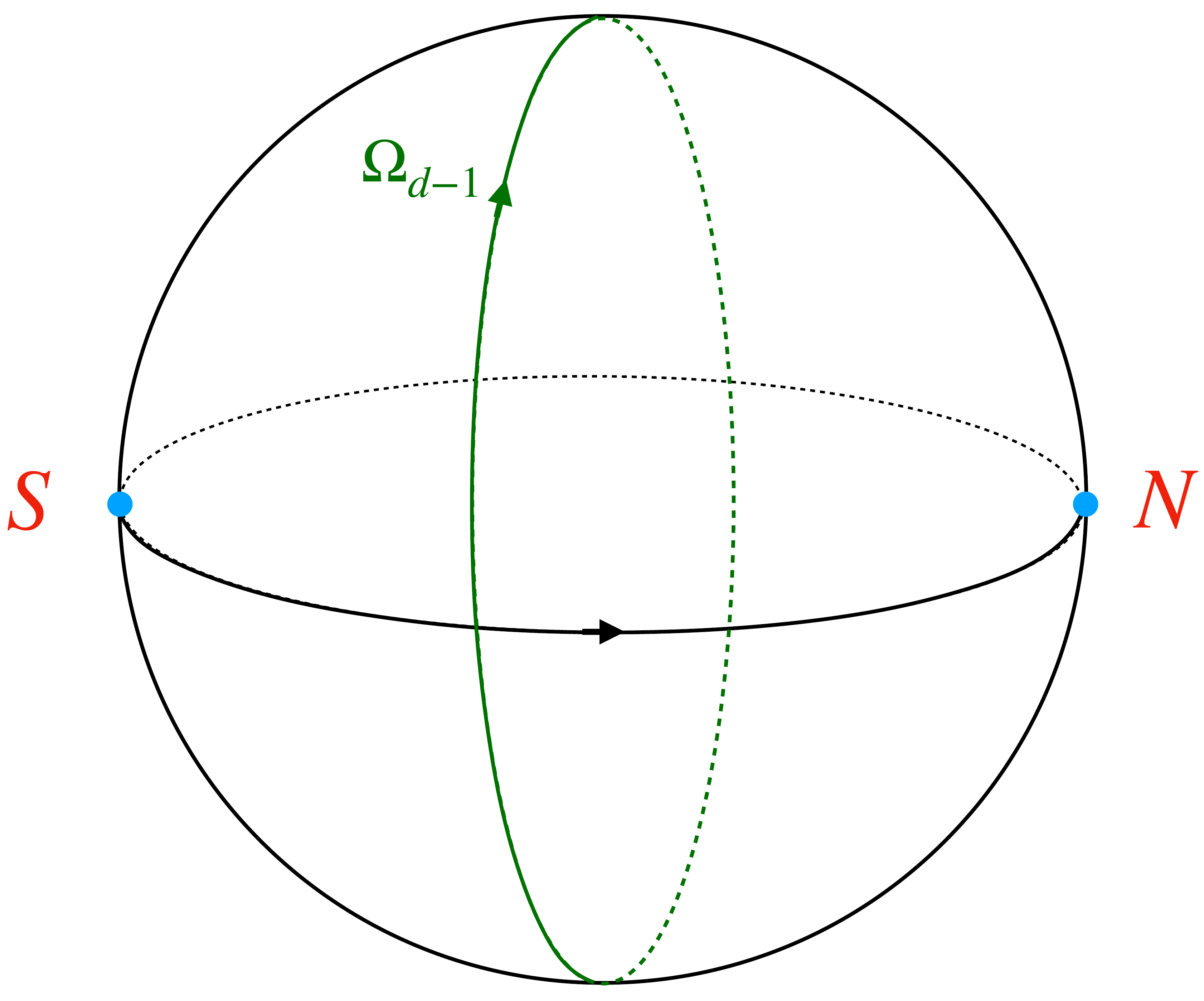}
  \caption{Left: dS$_{d+1}$/dS$_{d}$ slicing of $d+1$-dimensional de Sitter spacetime. The colorful curves denote $d$-dimensional dS branes with various tensions. Right: A time slice of global dS$_{d+1}$ spacetime that consists of two hemispheres glued by a zero tension brane.}
\label{fig:gluedS}
\end{figure}

One of our motivations is to construct holography without boundaries. Unlike AdS spacetime, which has a timelike conformal boundary, de Sitter spacetime is a naturally closed universe. It is straightforward to generalize our analysis to asymptotically de Sitter spacetime. When two $(d+1)$-dimensional dS vacuums are glued together by a timelike hypersurface, the Hamiltonian constraint on the brane is given by
\begin{equation}\label{eq:HamiltoniandS}
R= K^2 - K^{\mu\nu} K_{\mu\nu}  +\frac{d(d-1)}{L^2_{\mt{dS}}} \,, 
\end{equation}
which can also be derived from the AdS counterpart by performing the analytical continuation $L_{\mt{AdS}} \to iL_{\mt{dS}}$. 
The timelike brane living in de Sitter spacetime with a tension $T$ is thus constrained by a similar equation, \viz 
\begin{equation}
 R  +   \mu = - \( \langle  \tau^{(1)ij} \rangle \langle \tau^{(1)}_{ij} \rangle -\frac{\langle \tau^{(1)} \rangle^2   }{d-1}  \)=- \( \langle  \tau^{(2)ij} \rangle \langle \tau^{(2)}_{ij} \rangle -\frac{\langle \tau^{(2)} \rangle^2   }{d-1}  \) \,,
\end{equation}
but with identifying the Liouville potential as 
\begin{equation}
 \mu =  -\frac{d(d-1)}{L^2_{\mt{dS}_a}} -  \frac{d}{d-1}T^2 \le 0 \,. 
\end{equation}
Contrary to the AdS case, it is obvious that the brane in dS space is always associated with a positive curvature when the $T\bar{T}$ term vanishes due to the positivity of the potential term. The simplest examples of gluing two dS spacetimes can be found by considering the symmetric case where each side is given by a dS$_{d+1}$ spacetime with a dS$_{d}$ brane as the boundary (see \eg \cite{Geng:2021wcq}). It is nothing but the dS$_{d+1}$/dS$_{d}$ slicing as shown in Figure \ref{fig:gluedS}. It's worth noting that dS$_{d+1}$ spacetime can be thought of as a closed universe created by gluing two half dS$_{d+1}$ spacetimes along a $d$-dimensional brane whose tension vanishes. This also motivates us to consider constructing non-boundary AdS spacetime by gluing two AdS spacetimes together with a brane.

\subsection*{Mixed bulk geometries}

\begin{figure}[ht]
  \centering
   \includegraphics[width=3in]{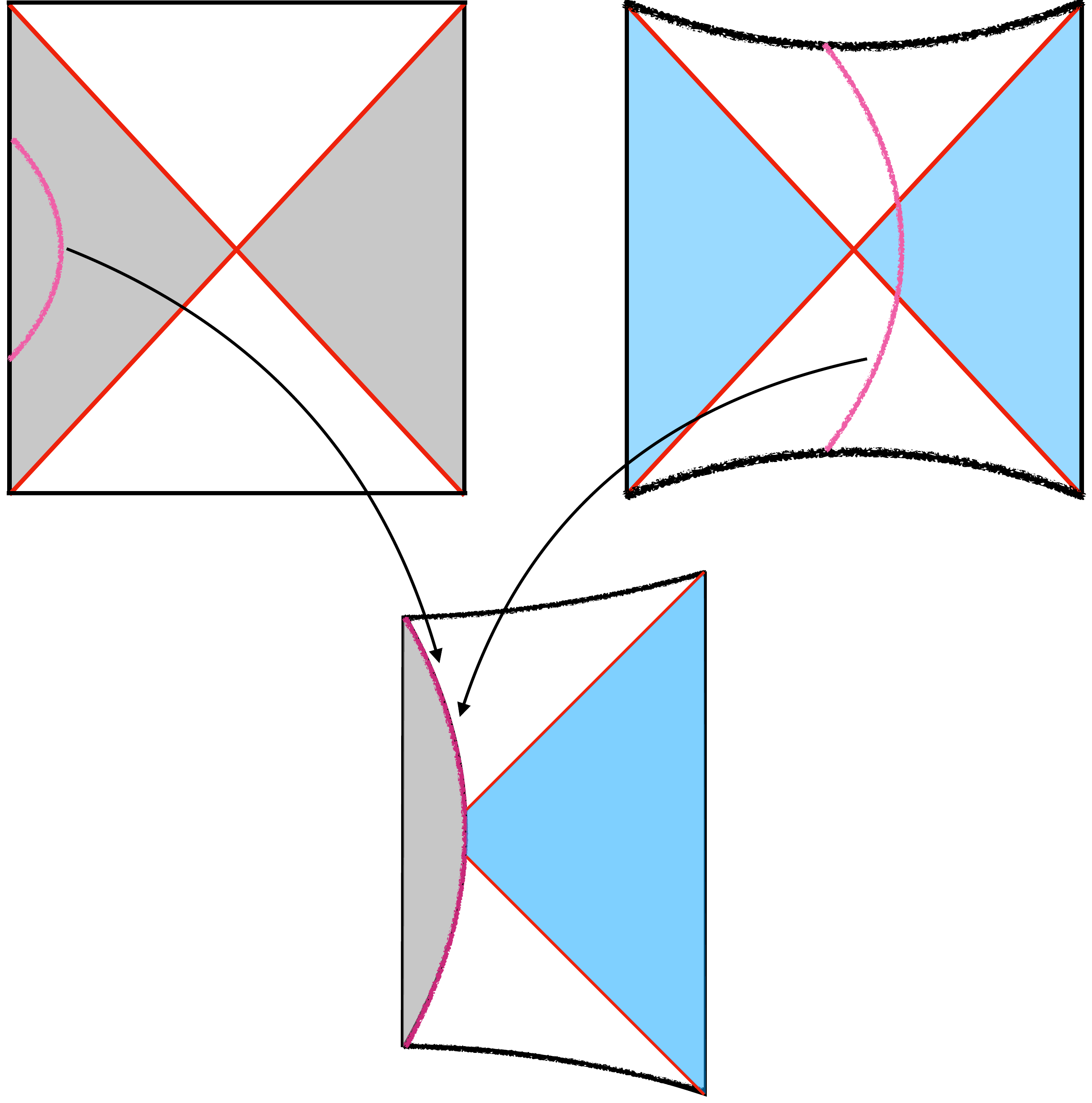}
  \caption{A sketch of one possible configuration for gluing dS spacetime with AdS-Schwarzschild spacetime along a timelike hypersurface. The trajectory of the brane is denoted by the pink curve.}
\label{fig:glueAdSdS}
\end{figure}

It is also straightforward to consider gluing two asymptotically flat spacetime. The constraint equations have a similar form, but with $\mu = - \frac{d}{d-1}T^2$, which can be derived from the AdS case by setting $L_{\mt{AdS}} \to \infty$. More generally, one can glue two different types of spacetimes. Let us consider two vacuum spacetimes in Einstein gravity with distinct cosmological constants $\Lambda^{(1)}, \Lambda^{(2)}$. The Israel junction condition still fixes the hypersurface $\Sigma^{(a)}$ with respect to each side as a CMC slice where the trace of the extrinsic curvature is given by 
 \begin{equation}\label{eq:Kconstant02}
		K^{(1)} = \frac{d\, T}{d-1} - \frac{\Lambda_1 - \Lambda_2}{2T} \,, \quad 		K^{(2)} = \frac{d \, T}{d-1} + \frac{\Lambda_1 - \Lambda_2}{2T}\,. 
	\end{equation}
 Similarly, we can rewrite the corresponding constraint equation in terms of the brane stress tensor $\tau_{ij}$, \viz 
 \begin{equation}
   R  +  \mu + \( \tau^{(a)ij} \,  \tau^{(a)}_{ij}-\frac{(\tau^{(a)})^2   }{d-1}  \)= -\frac{2 T^{(a)}}{d-1} \tau^{(a)} =0 \,,
 \end{equation}
with 
\begin{equation}
    T^{(a)}= \frac{d-1}{d} K^{(a)}\,, \qquad \mu= - \frac{d}{d-1} \( T^{(a)} \)^2 - 2 \Lambda_a\,. 
\end{equation}
By varying the cosmological constants $\Lambda^{(a)}$, one may construct six distinct types of joint spacetime, some of which have been studied before from different viewpoints. For example, see \cite{Blau:1986cw,Alberghi:1999kd,Aguirre:2005xs} for dS spacetime glued with asymptotically flat spacetime, and \eg \cite{Freivogel:2005qh,Fu:2019oyc,Lowe:2010np,Chapman:2021eyy,Auzzi:2023qbm} for dS spacetime glued with AdS-Schwarzschild spacetime as shown in Figure \ref{fig:glueAdSdS}.

%%%%%%%%%%%%%%%%%%%%%%%%%%%%%
%%%%%%%%%%%%%%%%%%%%%%%%%%%%%%
\section*{Acknowledgements}
%%%%%%%%%%%%%%%%%%%%%%%%%%%%%%%%%%%%%%%%%%%%%%%%%%%%%%%%
%%%%%%%%%%%%%%%%%%%%%%%%%%%%%%%%%%%%%%%%%%%%%%%%%%%%%%%%

We are grateful to Keisuke Izumi, Yuya Kusuki, Mukund Rangamani, and Zixia Wei for useful discussions. This work is supported by the Simons Foundation through the ``It from Qubit'' collaboration and by MEXT KAKENHI Grant-in-Aid for Transformative Research Areas (A) through the ``Extreme Universe'' collaboration: Grant Number 21H05187. TT is also supported by Inamori Research Institute for Science and by JSPS Grant-in-Aid for Scientific Research (A) No.~21H04469. SMR is also supported by JSPS KAKENHI Research Activity Start-up Grant Number JP22K20370.

\appendix

%%%%%%%%%%%%%%%%%%%%%%%%%%%%%%%%%%%%%%%%%%%%%%%%%%%%%%%%
%%%%%%%%%%%%%%%%%%%%%%%%%%%%%%%%%%%%%%%%%%%%%%%%%%%%%%%%
% bibliography via BibTeX
\bibliographystyle{jhep}
\bibliography{gluingAdS}

%%%%%%%%%%%%%%%%%%%%%%%%%%%%%%%%%%%%%%%%%%%%%%%%%%%%%%%%
%%%%%%%%%%%%%%%%%%%%%%%%%%%%%%%%%%%%%%%%%%%%%%%%%%%%%%%%

\end{document}